\documentclass[11pt]{article}
\usepackage[usenames,dvipsnames,svgnames,table]{xcolor} 
\usepackage[obeyspaces,hyphens,spaces]{url}
\usepackage{jcapmod}

\usepackage{booktabs}
\usepackage[english]{babel}
\usepackage{amsmath, amssymb, amsbsy, amstext, amsthm, simplewick}
\usepackage{hyperref}
\usepackage{graphicx}
\usepackage{amsfonts}
\usepackage{enumitem}
\usepackage{upgreek}
\usepackage{framed}
\usepackage{tensor}
\usepackage{pifont}
\usepackage{latexsym, mathrsfs}
\usepackage{array}
\usepackage{hyperref}
\usepackage{xspace}
\usepackage{longtable}
\usepackage{multirow}
\usepackage{cases}
\usepackage{empheq}
\usepackage{bm}
\usepackage{bbm}

\usepackage{braket}
\usepackage[utf8]{inputenc}

\usepackage[load-configurations=astronomy, range-units=brackets, range-phrase=-, per-mode=reciprocal, mode=math]{siunitx}
\usepackage{threeparttable}
\usepackage{subcaption}
\usepackage{graphicx}
\usepackage{ragged2e}

\usepackage{hhline}
\usepackage{array}
\newcolumntype{P}[1]{>{\centering\arraybackslash}p{#1}}
\usepackage{booktabs}

\definecolor{Blue}{rgb}{0.25, 0.41, 0.88}
\definecolor{Red}{rgb}{0.92,0.,0.}
\definecolor{darkorange}{rgb}{1.0,0.549,0.}
\definecolor{cobalt}{RGB}{44, 98, 120}
\definecolor{Mathematica1}{rgb}{0.368417, 0.506779, 0.709798}
\definecolor{Mathematica2}{rgb}{0.880722, 0.611041, 0.142051}
\definecolor{Mathematica3}{rgb}{0.560181, 0.691569, 0.194885}
\definecolor{Mathematica4}{rgb}{0.922526, 0.385626, 0.209179}
\definecolor{Mathematica5}{rgb}{0.528488, 0.470624, 0.701351}
\definecolor{Mathematica6}{rgb}{0.772079, 0.431554, 0.102387}
\definecolor{Mathematica7}{rgb}{0.363898, 0.618501, 0.782349}
\definecolor{Mathematica8}{rgb}{1, 0.75, 0}
\definecolor{Mathematica9}{rgb}{0.647624, 0.37816, 0.614037}
\definecolor{plotBlue}{RGB}{94, 130, 181}
\definecolor{plotRed}{RGB}{233, 85, 54}
\definecolor{plotGreen}{RGB}{142, 176, 50}
\definecolor{plotPurple}{RGB}{135, 120, 178}

\definecolor{cornellRed}{HTML}{B31B1B}
\definecolor{cornellBlue}{HTML}{0068AC}
\definecolor{cornellGreen}{HTML}{6EB43F}

\newcolumntype{C}[1]{>{\centering\let\newline\\\arraybackslash\hspace{0pt}}m{#1}}

\def\d{{\rm d}}

\makeatletter
\newlength{\apb@width}
\newcommand{\autoparbox}[2][c]{\settowidth{\apb@width}{#2}\parbox[#1]{\apb@width}{#2}}

\makeatother

\numberwithin{equation}{section}

\def\beq{\begin{equation}}
\def\eeq{\end{equation}}

\def\bea{\begin{eqnarray}}
\def\eea{\end{eqnarray}}

\newcommand{\lib}[1]{\texttt{#1}}

\def\d{{\rm d}}

\def\beq{\begin{equation}}
\def\eeq{\end{equation}}
\def\bea{\begin{eqnarray}}
\def\eea{\end{eqnarray}}

\def\d{{\rm d}}

\def\d{{\rm d}}

\DeclareRobustCommand{\SkipTocEntry}[4]{}

\usepackage{colortbl}
\definecolor{blue2}{cmyk}{1, 0.1, 0.1, 0}

\definecolor{pyBlue}{RGB}{31, 119, 180}
\definecolor{pyRed}{RGB}{214, 39, 40}
\definecolor{pyGreen}{RGB}{44, 160, 44}
\definecolor{pyBlue2}{RGB}{0, 111, 237}
\definecolor{pyRed2}{RGB}{224, 52, 36}

\makeatletter
\def\Ddots{\mathinner{\mkern1mu\raise\p@
\vbox{\kern7\p@\hbox{.}}\mkern2mu
\raise4\p@\hbox{.}\mkern2mu\raise7\p@\hbox{.}\mkern1mu}}
\makeatother

\begin{document}

\pagenumbering{roman}
\begin{titlepage}
\baselineskip=15.5pt \thispagestyle{empty}
\begin{flushright}

\end{flushright}
\vspace{-0.5cm}

\begin{center}
{\fontsize{19}{24}\selectfont  \bfseries Self-Interacting Gravitational Atoms \\[6pt] in the Strong-Gravity Regime}
\end{center}

\vspace{0.05cm}
\begin{center}
{\fontsize{12}{18}\selectfont Horng Sheng Chia$^{1}$, Christoffel Doorman$^{2,3}$, Alexandra Wernersson$^{3, 4}$, \\[2pt] Tanja Hinderer$^{4}$,  and Samaya Nissanke$^{3, 5}$} 
\end{center}

\begin{center}
\vskip8pt

\textsl{$^1$ School of Natural Sciences, Institute for Advanced Study, Princeton, NJ 08540, United States}

\vskip8pt

\textsl{$^2$ Department of Computer Science, University College London, \\ Gower Street, London WC1E 6BT, United Kingdom}

\vskip8pt

\textsl{$^3$ Gravitation Astroparticle Physics Amsterdam (GRAPPA), University of Amsterdam, \\ Science Park 904, Amsterdam, 1098 XH, The Netherlands}

\vskip8pt

\textsl{$^4$ Institute for Theoretical Physics, Utrecht University, Princetonplein 5, \\ 3584 CC Utrecht, The Netherlands}

\vskip8pt

\textsl{$^5$ Nikhef, Science Park 105, 1098 XG Amsterdam, The Netherlands}

\end{center}

\vspace{0.3cm}
\hrule \vspace{0.3cm}
\noindent {\bf Abstract}\\[0.1cm]
We numerically investigate free and self-interacting ultralight scalar fields around black holes in General Relativity. We focus on complex scalar fields $\Phi$ whose self-interactions are described by the quartic potential $V \propto \lambda |\Phi|^4$, and ignore the black hole spin in order to disentangle the effects of self interactions on the boson cloud. Using the spectral solver \texttt{Kadath}, we compute quasi-equilibrium configurations of the dominant eigenstates, including their backreaction on the spacetime metric. For scenarios with $- 10^{-2} \lesssim \lambda \lesssim 10^{-2}$ 
we find the mass of the self-interacting scalar cloud to be up to $\sim 70\%$ larger than that of a free scalar cloud, though the additional backreaction effect on the spacetime metric is only up to $\sim 1\%$ due to the low-density nature of the bosonic configurations. In this region of parameter space we observe approximate quadratic scalings between the mass of the cloud with $\lambda$, the scalar field amplitude, and the couplings between these two parameters. For systems with $\lambda$ beyond this range, the eigenfrequencies differ sufficiently from the known free-test-field values used as inputs in our numerical setup
to make the results, though convergent, physically unreliable.
This bounds the range of $\lambda$ in which the free scalar field solution remains a good approximation to self-interacting scalar field configurations. Our work is among the first nonperturbative explorations of self-interacting bosonic clouds around black holes, yielding detailed new insights into such systems in the nonlinear regime, while also overcoming technical challenges and quantifying limitations. Additionally, our results provide useful inputs
for fully dynamical numerical relativity simulations and for future explorations of spinning black holes and real scalar fields. 

\vskip10pt
\hrule
\vskip10pt

\end{titlepage}

\thispagestyle{empty}
\setcounter{page}{2}
\tableofcontents

\newpage
\pagenumbering{arabic}
\setcounter{page}{1}

\clearpage
\section{Introduction}
 \label{sec:introduction}

\vskip 4pt

After decades of theoretical and experimental explorations, a central mystery in physics remains unresolved: what is dark matter? Complementary to investigations into particle dark matter models, such as weakly-interacting massive particles, there is a growing interest in the wave dark matter paradigm in which dark matter consists of \textit{ultralight bosons}~\cite{Hu:2000ke, Hui2017, Hui:2021tkt, Schive2014, Harko2015, Ferreira2020}. These hypothetical boson fields have masses that could be much smaller than the sum of the neutrino masses; though they display coherent wave-like properties at scales smaller than their de Broglie wavelength, they behave essentially as cold dark matter at larger scales. These ultralight fields arise naturally in many theories beyond the Standard Model and are promising candidates to solve various open problems in contemporary physics~\cite{Peccei1977, Weinberg:1977ma, Wilczek:1977pj,Preskill:1982cy, Arvanitaki2010, Chadha-Day:2021szb, Hook:2018dlk}. In fact, their conjectured masses span numerous orders of magnitude, with typical scenarios in string theory compactifications predicting the presence of a plethora of such bosons~\cite{Svrcek:2006yi, Baumann:2014nda, Cicoli:2021gss}. As these bosons are expected to couple extremely weakly with ordinary matter, they could have easily been missed in experimental searches to date~\cite{Jaeckel2010, Essig2013, Berlin:2018bsc}. 

\vskip 4pt

By virtue of the equivalence principle, gravity is the only portal through which these ultralight bosons are guaranteed to interact with the Standard Model, making gravitational probes of these boson fields promising avenues to test for such beyond the Standard Model physics. Remarkably, this minimal coupling between these ultralight bosons and rotating black holes, which are the cleanest and most strongly gravitationally bounded bodies, would naturally induce an instability known as \textit{black hole superradiance}~\cite{Zeldovich1972, Starobinsky:1973aij, Bekenstein:1973mi, Brito2015}. Black hole rotational superradiance is an amplification process in which energy and angular momentum are extracted from the black hole triggered when the the black hole angular frequency $\Omega_H$ satisfies
\begin{equation}
    \Omega_H > \frac{\omega}{m} \, \label{eqn:inequality}
\end{equation}
where $\omega$ is the eigenfrequency of the boson field and $m$ is its azimuthal angular momentum. As energy and angular momentum are continuously extracted from the black hole during the amplification process, $\Omega_H$ would decrease until the inequality (\ref{eqn:inequality}) is saturated and the process halts. The amount of energy extracted from a Kerr black hole through the superradiance process can be enormous, up to the Penrose limit of $29\%$ of the initial black hole mass~\cite{Penrose:1969pc, Hawking1973} . 

\vskip 4pt

For astrophysical black holes, whose Schwarzschild radii range between $[1, 10^{10}]$ km, the superradiant instability is most efficiently triggered by ultralight bosons with masses $\left[10^{-20}, 10^{-10}\right]$ eV, making astrophysical black holes interesting laboratories of physics beyond the Standard Model~\cite{Arvanitaki2011, Baryakhtar:2017ngi, Baumann2019, Mehta:2021pwf}. Remarkably, the resulting structure of the cloud has many similarities to the electron energy levels in the hydrogen atom, with the role of the fine structure constant $\alpha$ played by the ratio of the Compton wavelength of the field to the radius of the black hole. However, due to the non-axisymmetric and non-stationary nature of these boson clouds, they emit gravitational radiation and are not long-term stable states~\cite{Arvanitaki2011, Yoshino2014, Arvanitaki2015, Brito:2017wnc}. In addition, boson clouds with non-trivial scalar self interactions exhibit a plethora of interesting scalar wave signatures, such as bosenova explosions and coherent scalar wave emissions~\cite{Arvanitaki2011, Yoshino:2012kn, Baryakhtar:2020gao, Omiya:2022gwu}, which could be detectable by table-top axion experiments~\cite{Kahn:2016aff, JacksonKimball:2017elr, Chaudhuri:2019ntz, Lasenby:2019prg, Berlin:2020vrk}. Furthermore, in merging black hole binary system, the presence of a cloud is expected to significantly affect the dynamics of binary systems, for instance leading to non-perturbative resonant transitions between eigenstates or possibly tidal disruption~\cite{Chia:2020dye, Baumann2020a, Zhang:2018kib, Baumann2019a, Ding:2020bnl, Baumann:2021fkf}, resulting in distinctive imprints in the gravitational waveforms that could be measurable by current gravitational wave observatories~\cite{LIGOScientific:2014pky, VIRGO:2014yos,KAGRA:2020tym} or future detectors~\cite{LISA:2022kgy,Kalogera:2021bya,Maggiore:2019uih}. To look for these gravitational and scalar wave signatures from ultralight bosons, robustly interpret the recent and upcoming detections and place constraints on the axion parameter space requires a detailed understanding of these bosonic configurations and their interplay with strong-field gravity. 

\vskip 4pt

Studies over the past decade have elucidated a wealth of interesting features of scalar bosonic configurations around black holes, though they also had limitations. For instance: 

\begin{itemize}
\item \textit{Scalar self interactions} -- While bosonic clouds around black holes for free scalar fields have been investigated extensively in the literature~\cite{Detweiler1980, Dolan:2007mj,  Arvanitaki2011, Yoshino2014, Herdeiro:2014goa, Brito2015a,  Brito:2017wnc, Baumann2019}, explorations into solutions of self-interacting scalar fields have been relatively limited~\cite{Arvanitaki2011, Yoshino:2012kn, Baryakhtar:2020gao, Omiya:2022gwu}. The reason is that the nonlinearities arising from a scalar field potential introduce significant technical challenges, both analytical and numerical, which prevent a comprehensive exploration. Non-trivial scalar potentials, such as a periodic potential arising from non-perturbative instanton effects in string axions or a quartic self interaction operator in the scalar field effective field theory, arise naturally from the perspective of more fundamental, ultraviolet (UV) complete theories and are almost bound to arise in realistic scenarios for these ultralight scalar fields;

\item \textit{Backreaction onto the metric} -- Most studies conduct a test field analysis, in which the background spacetime is fixed to be the Kerr geometry and the gravitational backreaction of the bosonic configuration onto the metric is assumed to be negligible~\cite{Detweiler1980, Dolan:2007mj, Brito2015a, Baumann2019}. Given the low-density nature of these boson clouds, this assumption is certainly justified, though a precise quantification of the magnitude of backreaction effects would be desirable (see e.g.~\cite{Okawa:2014nda, East:2017ovw, East:2017mrj, Wang:2022hra} for full relativistic backreaction effects of boson field configurations onto the backgroud spacetime, mostly for massive vector fields because the instability timescales are much shorter than those for scalars).

\end{itemize}

\vskip 4pt

In this work, we perform numerical computations for bosonic configurations of free and self-interacting complex scalar fields with a quartic self-interaction potential $V \propto \lambda |\Phi|^4$ around black holes in General Relativity. We use the spectral code \texttt{Kadath}~\cite{Grandclement2010} to compute the nonlinear solutions of the boson clouds and the spacetime metric from the Einstein field equations coupled with the Klein-Gordon equation of motion in a quasi-equilibrium approximation that focuses on the state of the system after the cloud has stopped growing. Specifically, our approach involves solving the Hamiltonian and momentum constraints in the $3+1$ decomposition~\cite{Arnowitt:1959ah} of the Einstein-Klein-Gordon field equations, supplementing them  with the trace of the spatial evolution equation in what is known as the \textit{extended conformal thin sandwich (XCTS) method}~\cite{Pfeiffer2003, Gourgoulhon:2007ue}. This method is regularly adopted in solving for the initial data in dynamical numerical relativity simulations and, unlike canonical approaches to solving the initial data, more accurately captures the physics in the numeric implementation. Indeed, we shall see how our results provide quantitative insights into the effects of $\lambda$ on the scalar field properties and the degrees to which gravitational backreaction of the clouds on the metric are important in an approximate yet more controlled setting than full dynamical evolutions. 

\vskip 4pt

Although light real scalar fields arise more naturally from UV models, for instance, Goldstone modes of spontaneously broken global symmetries and axions or axion-like particles~\cite{Peccei1977, Arvanitaki2010}, we focus here on complex scalar fields because the $U(1)$ symmetry of complex fields enforces stationarity and axisymmetry. Conversely, the bosonic configurations of real scalar fields are non-stationary and non-axisymmetric, which would introduce additional technical complications to our numeric setup. It is worth pointing out that complex scalar fields with a quartic potential $V \propto \lambda |\Phi|^4$ are often discussed in the context of boson stars~\cite{Kaup:1968zz, Ruffini:1969qy, Breit:1983nr, Colpi:1986ye, Eby:2015hsq, Liebling:2012fv, Visinelli:2021uve}. We emphasize however that boson stars belong to an entirely different class of self-gravitating scalar field solutions: not only would they form through gravitational collapse or through phase transitions in the early universe, but they also have regular boundary conditions at their centers. By contrast, the boson clouds studied in this work would  spontaneously form around rotating black holes and have black holes at their centers~\cite{Liebling:2012fv, Visinelli:2021uve}. In fact, we shall describe how a major technical challenge in this work involves a careful treatment of the singularity structure of the Klein-Gordon equation at the black hole event horizon, which in turn determines many key properties of the boson clouds. 

\vskip 4pt 

To cleanly separate any potential degeneracy between the black hole spin and scalar self interactions, we simplify our analysis by restricting ourselves to a spherically symmetric spacetimes. Although black hole spin is a necessary ingredient in the natural formation of these bosonic configurations, the final black hole spin after the saturation of the superradiant inequality (\ref{eqn:inequality}) is often small and its gravitational influence on the cloud is subdominant compared to that of the black hole mass. Owing to this simplified setup, the bulk of our analysis will be focused on the gauge-invariant ADM mass of the entire system as a measure of its backreaction effecs of the cloud on the metric. Our work offers detailed understanding of self-interacting boson clouds in General Relativity and provides a useful first step towards future explorations of these systems.

\paragraph{Outline}  The plan of the paper is as follows: in Section~\ref{sec:setup}, we describe our numerical methods to solve the boson cloud as an initial value problem in General Relativity. We will also describe technical aspects of the numerical evaluation that are unique to the gravitational atom. In Section~\ref{sec:numerics}, we present our numerical results and quantify the nonlinearities of our solutions. Finally, we conclude and provide an outlook in Section~\ref{sec:conclusion}.

\paragraph{Notations and Conventions} Greek indices $\mu, \nu, \cdots$ denote spacetime indices while Latin indices $i, j, \cdots $ denote spatial indices. We work in natural units, $G = c = \hbar = 1$. Our choices of boundary conditions for the background metric, i.e. the spacetime without taking into account the backreaction of the cloud, generates a solution that is equivalent to the Schwarzschild black hole in isotropic coordinates:
\begin{equation}
    \d s^2=-\left(\frac{2r-M}{2r+M} \right)^2 \d t^2 + \left(1 + \frac{M}{2r} \right)^4
    \left(\d r^2+r^2 \hskip 1pt \d \theta^2 + r^2 \sin^2 \theta \hskip 1pt \d \phi^2 \right) \, ,
\end{equation}
where $M$ is the black hole mass. In these coordinates, the event horizon is located at $r = r_{\rm BH} \equiv M/2$. In our numerical domain, $r_{\rm BH}$ defines the unit radius of the domain size. We emphasize that $M$ is to be distinguished from the ADM mass of the total system, $M_{\rm ADM}$, which includes the mass of the boson cloud when the backreaction of the cloud on the metric is taken into account. The mass of the cloud is therefore defined as $M_c = M_{\rm ADM} - M$.

\vskip 4pt

The lapse function and conformal factor evaluated on the Schwarzschild background spacetime are denoted by $\overline{N}$ and $\overline{\Psi}$ in order to distinguish them from their fully nonlinear gravitational counterparts, $N$ and $\Psi$. For the scalar field $\Phi$, we also introduce the normalized field, $\Phi^{N}$, for ease of comparison between scalar field profiles with different choices of amplitudes.

\vskip 4pt

We use the terms ``gravitational atom" and ``boson clouds" interchangeably for scalar field configurations around black holes. For the free scalar field, we label the eigenstates by $\ket{n \ell m}$, where $n \geq \ell + 1$ is the principal quantum number, $\ell$ is the orbital angular momentum number and $m$ is the azimuthal quantum number. For self-interacting fields, we label the states with the additional parameter, $\lambda$. The gravitational fine structure constant is $\alpha = M \mu$, where $\mu$ is the boson mass. Crucially, $\alpha$ is defined with respect to $M$ but not $M_{\rm ADM}$.

\pagebreak

\section{Setup for the Initial Value Problem} \label{sec:setup}

In this section, we discuss our numerical methods in solving the gravitational atom as an initial value problem in General Relativity. We review the equations of a general Einstein-Klein-Gordon system and describe their formulations in the extended conformal thin sandwich method~\cite{Pfeiffer2003, Gourgoulhon:2007ue} in \S\ref{sec:EOM}.  We then specialize to the gravitational atom, discussing technical aspects of their numerical computations in \S\ref{sec:Solve}. 

\subsection{The Einstein and Klein-Gordon Equations} \label{sec:EOM}

We obtain numerical solutions of the self-interacting gravitational atom by solving the Einstein field equation coupled to the Klein-Gordon equation. To enforce stationary and axisymmetry, which reduces challenges of our numerical setup, we focus on field configurations of a complex scalar field, which respects a global $U(1)$ symmetry.
The equations are
\begin{align}
 & R_{\mu\nu} - \frac{1}{2} Rg_{\mu\nu}  = 8\pi T_{\mu\nu}\, , \label{eq:EE} \\
    & g^{\mu\nu} \nabla_{\mu}\nabla_{\nu}\Phi - \frac{\d V}{\d |\Phi|^2} \Phi  = 0 \, , \label{eqn:KG} 
\end{align}
where $g_{\mu \nu}$ is the metric, $R_{\mu \nu}$ is the Ricci tensor, $R \equiv g^{\mu \nu} R_{\mu \nu} $ is the Ricci scalar, $T_{\mu \nu}$ is the energy-momentum tensor of the matter field, and $V$ is the scalar field potential. In addition, since $\Phi$ is complex, the full system of equations also includes the complex conjugate of (\ref{eqn:KG}), which we omit here for brevity. In this work, we consider the quartic potential
\begin{equation}
    V (|\Phi|^2) = \mu^2 |\Phi|^2 + \frac{\lambda}{2} |\Phi|^4 \, , \label{eqn:phi^4} 
\end{equation}
where $\mu$ is the mass of the boson, and $\lambda$ is the self-interaction parameter. The energy momentum tensor of the scalar field with the potential (\ref{eqn:phi^4}) is
\begin{equation} \label{eqn:EMtensor}
    T_{\mu\nu}=\nabla_\mu\Phi\nabla_\nu\Phi^*+\nabla_\nu\Phi\nabla_\mu\Phi^*-g_{\mu\nu}\left(g^{\alpha\beta}\nabla_\alpha\Phi\nabla_\beta\Phi^*+\mu^2|\Phi|^2+\frac{\lambda}{2}|\Phi|^4\right).
\end{equation}
In this work, we investigate solutions of boson clouds around black holes in the free-field ($\lambda=0$), attractive ($\lambda < 0$), and repulsive ($\lambda > 0$) cases. Crucially, the non-linear gravitational interaction between the scalar field and the metric are included in all of these scenarios.

\subsubsection{$3+1$ Decomposition} \label{sec:31_decomposition}

Following the Arnowitt-Deser-Misner (ADM) formalism of General Relativity~\cite{Arnowitt:1959ah}, we foliate spacetime through the $3+1$ decomposition in the $\{t, x^i\}$ coordinates 
\begin{equation}\label{eq:3+1_line_element}
    \d s^2 = -N^2 \d t^2 + \gamma_{ij} (\d x^i + \beta^i \d t)(dx^j + \beta^j \d t)\,.
\end{equation}
where $N$ is the lapse function, $\beta^i$ is the shift vector, and $\gamma_{ij}$ is the induced spatial 3-metric. In the ADM formalism, the second-order partial differential equation (\ref{eq:EE}) is decomposed into a set of coupled first-order partial differential equations by introducing the extrinsic curvature of the spatial hypersurfaces,
\begin{equation}
    K_{ij} = - \frac{1}{2} \mathcal{L}_n \gamma_{ij} = - \frac{1}{2N} (\partial_t \gamma_{ij} - D_i \beta_j - D_j \beta_i) \, , \label{eqn:Kij}
\end{equation}
where $\mathcal{L}_n$ is the Lie derivative with respect to the hypersurface normal 
\begin{equation}
    n_\mu = (-N, 0, 0, 0),
\end{equation} $\partial_t$ is the partial derivative with respect to $t$, and $D_i$ is the covariant derivative with respect to $\gamma_{ij}$.

\vskip 4pt

Projecting (\ref{eq:EE}) onto the time-time and time-space components in the 3+1 decomposition leads to the Hamiltonian and momentum constraints
\begin{align}
    {}^{(3)}R + K^2 - K_{ij}K^{ij} & = 16\pi\rho \, ,  \label{eqn:HamConst} \\
    D_j {K^j}_i - D_i K & = 8 \pi j_i \, , \label{eqn:MomConst}
\end{align}
where ${}^{(3)}R$ is the Ricci scalar of the induced metric, $K \equiv \gamma^{ij} K_{ij}$ is the trace of the extrinsic curvature,  $\rho \equiv n^\mu n^\nu T_{\mu \nu}$ is the matter energy density, and $j^i = -n^{\mu}\gamma^{\nu i}T_{\mu\nu}$ is the momentum density. Since (\ref{eqn:HamConst}) and (\ref{eqn:MomConst}) are independent of time, they must be satisfied at all spatial hypersurfaces, including the initial slice. Furthermore, since these equations are elliptical in nature, their solutions depend sensitively on the global behaviour on the entire foliation. This sensitivity presents important challenges to solving the initial value problem, and we describe ways of tackling these issues in \S\ref{sec:Solve}.

\vskip 4pt

To solve an initial value problem in General Relativity, it usually suffices to focus only on the constraint equations (\ref{eqn:HamConst}) and (\ref{eqn:MomConst}). However, as we shall see in~\S\ref{sec:XCTS}, the trace of the ADM dynamical evolution equation will serve as a useful supplementary condition to our system of equations. Projecting (\ref{eq:EE}) onto the space-space component and taking its trace leads to
\begin{equation}\label{eq:trace_evolution}
\begin{aligned}
    (\partial_t - \beta^i D_i)K = -D_i D^i N + N\Big(4\pi(\rho+S) + K_{ij}K^{ij}\Big)\,,
\end{aligned}
\end{equation}
where $S = \gamma^{ij} S_{ij} = \gamma^{ij}  \gamma^{\mu}_{\ i}\gamma^{\nu}_{\ j}T_{\mu\nu}$ is trace of the spatial stress tensor. In the non-relativistic limit, (\ref{eq:trace_evolution}) reduces to the Poisson equation.

\vskip 4pt

Similar to the Einstein equations, the Klein-Gordon equation (\ref{eqn:KG}) can be decomposed into a set of coupled first-order differential equations in the $3+1$ formalism by introducing the canonical momentum
\begin{equation}
    \Pi = - \frac{1}{2} \mathcal{L}_n \Phi = - \frac{1}{2N} \left( \partial_t \Phi - \beta^i D_i  \Phi \right) \, . \label{eqn:canonicalmom}
\end{equation}
As a result, (\ref{eqn:KG}) can be rewritten as
\begin{equation}
 \left( \partial_t - \beta^i D_i \right) \Pi = \frac{N}{2} \left[ - \gamma^{ij} D_i D_j \Phi + 2 K \Pi + \mu^2 \Phi + \lambda |\Phi|^2 \Phi \right] - \frac{1}{2}\gamma^{ij} D_i \Phi D_j N \, . \label{eqn:KG3+1}
\end{equation}
Since $\Phi$ is a complex scalar field, the scalar sector consists of doubling the set of equations above by including the complex conjugation of (\ref{eqn:canonicalmom}) and (\ref{eqn:KG3+1}).

\subsubsection{Extended Conformal Thin Sandwich Method} \label{sec:XCTS}

To solve the Hamiltonian~(\ref{eqn:HamConst}) and momentum~(\ref{eqn:MomConst}) constraints on an initial hypersurface, we transform $\gamma_{ij}$ and $K_{ij}$ into a set of variables that are more amenable to numerical evaluations~\cite{Bruhat1971, York1972Mapping, Pfeiffer2003, Gourgoulhon2007}. In particular, the conformal metric, $\tilde{\gamma}_{ij}$, is introduced through the transformation
\begin{equation}\label{eq:conf}
    \tilde{\gamma}_{ij} = \Psi^{-4} \hskip 1pt \gamma_{ij} \, , \qquad  \tilde{\gamma}^{ij}  = \Psi^{4} \hskip 1pt \gamma^{ij} \, ,
\end{equation}
where $\Psi$ is the conformal factor. Furthermore, we decompose the extrinsic curvature into its traceless component, $A_{ij} = K_{ij} - \frac{1}{3}K\gamma_{ij}$, and similarly define the conformal traceless extrinsic curvature,
\begin{equation}\label{eq:confA}
    \tilde{A}_{ij} = \Psi^{-4} A_{ij} \,, \qquad \tilde{A}^{ij} = \Psi^{4} A^{ij} \, .
\end{equation}
From (\ref{eq:conf}) and (\ref{eq:confA}), the definition of the extrinsic curvature (\ref{eqn:Kij}) can be rewritten in terms of the conformal quantities as
\begin{align}
    \tilde{A}^{ij} = \frac{1}{2N} \left( \partial_t \tilde{\gamma}^{ij} +  \tilde{D}^i \beta^j + \tilde{D}^j \beta^i - \frac{2}{3} \tilde{D}_k \beta^k \tilde{\gamma}^{ij} \right) \, , \label{eqn:Atilde}
\end{align}
where $\tilde{D}_i$ is the covariant derivative with respect to $\tilde{\gamma}_{ij}$. The Hamiltonian and momentum constraints written in terms of the conformal variables are 
\begin{align}
    & \tilde{D}_i \tilde{D}^i \Psi - \frac{1}{8} {}^{(3)}\hskip -1pt \tilde{R} \hskip 1pt \Psi + \left( \frac{1 }{8} \tilde{A}_{ij} \tilde{A}^{ij} - \frac{1}{12} K^2 \right) \Psi^5 = -  2 \pi \Psi^5  \rho \, , \label{eqn:ham_tilde}\\
    & \tilde{D}_j \tilde{A}^{ij} + 6 \tilde{A}^{ij} \tilde{D}_j \ln \Psi - \frac{2}{3} \tilde{D}^i K = 8 \pi \Psi^4 j^i \, , \label{eqn:mom_tilde}
\end{align}
where ${}^{(3)} \hskip -1pt \tilde{R}$ is the conformal three-dimensional Ricci scalar. From (\ref{eqn:Atilde}), (\ref{eqn:ham_tilde}) and (\ref{eqn:mom_tilde}) we see that the initial value problem has been recast into a set of equations where $\Psi$ and $\beta^i$ are solved by specifying the initial data for $\{ \tilde{\gamma}_{ij}, \partial_t \tilde{\gamma}_{ij}, K, N, \rho, j^i \}$ on the hypersurface.

\vskip 4pt

Although the equations above are relatively straightforward to solve, they require specifying initial data for $N$, which \textit{a priori} has no natural choice for a general physical setup. To circumvent this shortcoming, we adopt the \textit{extended conformal thin sandwich (XCTS) method}~\cite{Pfeiffer2003}, where those equations are supplemented with the trace of the evolution equation (\ref{eq:trace_evolution}) which solves for $N$. Expressed in terms of the conformal variables, (\ref{eq:trace_evolution}) is 
\begin{equation}\label{eq:XCTS_full}
    (\partial_t - \beta^i \tilde{D}_i ) K  = - \Psi^{-4} \left( \tilde{D}_i \tilde{D}^i N + 2 \tilde{D}_i N \tilde{D}^i \ln{\Psi} \right) + N \left( 4\pi \left( \rho + S \right) + \tilde{A}_{ij} \tilde{A}^{ij} + \tfrac{1}{3}K^2 \right) \, .
\end{equation}
In the XCTS formalism, we specify the initial data for $\{ \tilde{\gamma}_{ij}, \partial_t \tilde{\gamma}_{ij}, K, \partial_t K, \rho, j^i, S \}$ and solve for $\Psi, N,$ and $\beta^i$ through the system of equations (\ref{eqn:Atilde}), (\ref{eqn:ham_tilde}), (\ref{eqn:mom_tilde}), and (\ref{eq:XCTS_full}). While solving for (\ref{eq:XCTS_full}) comes at the expense of additional computational cost, it has the great benefit of providing a more natural and physically-interpretable prescription for the initial data.

\vskip 4pt

In this work, we adopt common choices for $\{ \tilde{\gamma}_{ij}, \partial_t \tilde{\gamma}_{ij}, K, \partial_t K \}$ that significantly simplify the set of equations above: \textit{i)} we choose a conformally-flat spatial hypersurface, $\tilde{\gamma}_{ij} = \delta_{ij}$, which also implies ${}^{(3)}\hskip -1pt \tilde{R} = 0$; \textit{ii)} we foliate spacetime by setting $K=0$, which is commonly known as the maximal slicing ~\cite{Gourgoulhon:2007ue}; and \textit{iii)} we only seek to obtain stationary solutions, which allows us to take $\partial_t \tilde{\gamma}_{ij} = \partial_t K = 0$ (note however that time derivatives acting on $\Phi$ do not vanish for stationary configurations of the scalar field, c.f. \S\ref{sec:stationaryfield}). The energy density, momentum density, and trace of the spatial stress tensor of the matter-field variables in this decomposition are given by
\begin{equation}
\begin{aligned} \label{eqn:mattervars}
    \rho & = 4\Pi \hskip 1pt \Pi^*+\mu^2|\Phi|^2+\frac{\lambda}{2}|\Phi|^4+\Psi^{-4}\Tilde{D}_i\Phi\Tilde{D}^i\Phi^* \, , \\
     j^i & = 2\Psi^{-4}\left(\Pi^*\tilde{D}^i\Phi+\Pi\tilde{D}^i\Phi^*\right) \, , \\
    S & = 12\Pi \hskip 1pt \Pi^*-\Psi^{-4}\Tilde{D}_i\Phi\Tilde{D}^i\Phi^*-3\mu^2|\Phi|^2-\frac{3}{2}\lambda|\Phi|^4. \, 
\end{aligned}
\end{equation}
We specify the boundary conditions on the fields $\Phi$ and $\Phi^*$ as explained in \S\ref{sec:bc} and solve for these fields via the Klein-Gordon equation, which in the conformal variables reads
\begin{equation}\label{eq:conformal_KG}
    \frac{\Psi^{-4}}{N}\tilde{D}_iN \tilde{D}^i\Phi + \Psi^{-4}\tilde{D}_i\tilde{D}^i \Phi + 2\Psi^{-5}\tilde{D}_i \Phi\tilde{D}^i\Psi + 2\mathcal{L}_n\Pi - 2K\Pi - \mu^2 \Phi - \lambda |\Phi|^2\Phi  = 0  \, ,
\end{equation}
together with the complex conjugate of ~\ref{eq:conformal_KG}. In \S~\ref{sec:Solve}, we will explore the types of boundary conditions and physical inputs that define the gravitational atom in $\Phi$. We will also discuss several technical aspects on the numerical implementation of (\ref{eq:conformal_KG}) that is unique to the gravitational atom. For the numerical evaluations, we find it convenient to decompose the complex fields $\Phi$ and $\Phi^*$ into two real scalar fields $\Phi_1$ and $\Phi_2$ via 
\begin{equation}
\label{eq:Phitoreal}
    \Phi = (\Phi_1 + i \Phi_2)/\sqrt{2} \, ,
\end{equation}
and $ \Phi^* = (\Phi_1 - i \Phi_2)/\sqrt{2}$.

\subsection{Solving for the Gravitational Atom} \label{sec:Solve}

The system of equations described in \S\ref{sec:EOM} applies to a general scalar field configuration in General Relativity. In this section, we solve those equations specifically for the gravitational atom. We describe technical aspects of our numerical implementation, many of which are unique to the gravitational atom and are different from other bosonic configurations, e.g. boson stars~\cite{Kaup:1968zz, Ruffini:1969qy, Breit:1983nr, Colpi:1986ye, Eby:2015hsq, Liebling:2012fv, Visinelli:2021uve}. A schematic illustration of the dominant atomic state for the free scalar field and the setup of our numerical domain are shown in Fig.~\ref{fig:domain}. The size and resolution of the regions in the domain are detailed in Table~\ref{table:resolution}, with convergence tests that led to our choices of resolution discussed in \S\ref{sec:redshiftpotential_Schw}.

\subsubsection{Boundary Conditions} \label{sec:bc}

The gravitational atom differs from other bosonic configurations in General Relativity due to the unique boundary condition near its origin. While configurations such as boson stars have regular boundary conditions at their origins~\cite{Liebling:2012fv, Visinelli:2021uve}, the gravitational atom is defined by the ``only-ingoing" boundary condition of the central black hole's event horizon. As we shall see, though this boundary condition is analytically well understood~\cite{Detweiler1980, Baumann2019}, it is non-trivial to implement numerically within the setup described above.

\vskip 4pt

In the XCTS formalism, we solve for $ \{ \Psi, N, \beta^i, \Phi \}$ by adopting common choices for other variables (see~\S\ref{sec:XCTS} for a detailed discussion). This requires specifying the boundary conditions for these quantities at the black hole horizon and at asymptotic infinity, which coincides with the outermost boundary of our numerical domain, $r = r_{\rm inf}$, due to the use of compactified coordinate. For the metric quantities $ \{ \Psi, N, \beta^i \}$, we impose the asymptotic flatness boundary conditions at $r = r_{\rm inf}$, which are
\begin{equation}
\begin{aligned}
\Psi \rvert_{r_{\rm inf}} = 1 \, , \qquad
    N \rvert_{r_{\rm inf}} = 1 \, , \qquad
    \beta^i \rvert_{r_{\rm inf} } = 0 \, . \label{eqn:bc_metric_inf}
\end{aligned}
\end{equation}
It is easy to see from (\ref{eq:3+1_line_element}) and (\ref{eq:conf}) and these choices recover the Minkowski spacetime at large distances. On the other hand, at the event horizon, $r=r_{\rm BH}$, we impose the following conditions for a Kerr black hole~\cite{Grandclement2010}
\begin{equation} 
    \begin{aligned}
        \left( \tilde{D}_r\Psi + \frac{\Psi}{2r} \right) \Big\rvert_{r_{\rm BH}} = -\frac{\Psi^{3}}{4}\tilde{A}_{ij} \hat{r}^i \hat{r}^j \, , \quad 
        N\rvert_{r_{\rm BH}} = n_{0} \, , \quad
        \beta^i\rvert_{r_{\rm BH}} &= n_{0}\Psi^{-2} \hskip 1pt \hat{r}^i + \Omega_H \hskip 1pt \hat{s}^i \, , \label{eq:bc_Kadath_paper}
    \end{aligned}
\end{equation}
where $\hat{r}^i \propto (x, y, z)$ is the unit vector centered on the sphere in Cartesian coordinates, $\Omega_H$ is the black hole angular velocity, $\hat{s}^i \propto (-y, x, 0)$ is the black hole spin axis and is orthonormal to $\hat{r}^i$, and the constant $n_0\geq 0$ determines the coordinates of our setup (see \ref{sec:Schwarzschild} below for further discussion). The boundary condition for $\Psi$ in (\ref{eq:bc_Kadath_paper}) is derived by demanding that the $r = r_{\rm BH}$ surface be a non-expanding horizon, i.e. the expansion of the null geodesic congruence of this horizon vanishes~\cite{Gourgoulhon:2005ng}. Although this condition does not hold for null hypersurfaces in dynamical environments, as the Raychaudhuri equation would dictate~\cite{hawking_ellis_1973}, it applies to black holes in equilibrium, which this work focuses on. The boundary condition for $\beta^i$ in (\ref{eq:bc_Kadath_paper}) enforces a coordinate system that is stationary with respect to the black hole horizon~\cite{Gourgoulhon:2005ng}.

\begin{figure}[t!]
    \centering
    \begin{subfigure}{0.45\textwidth}
    \vspace{2.5cm}
    \raisebox{0pt}{\includegraphics[trim=20 0 0 250, scale=1]{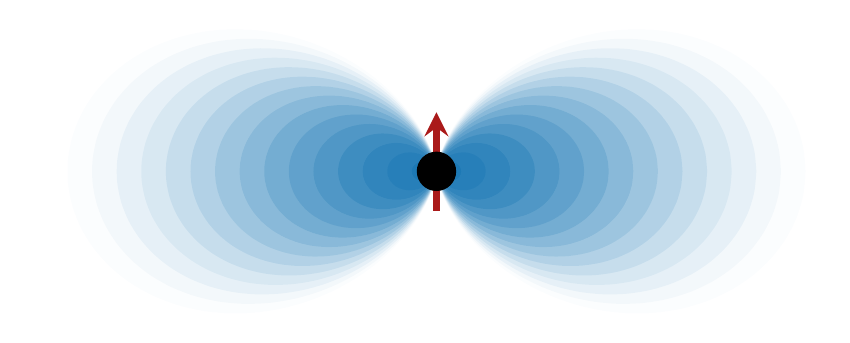}}
    \end{subfigure}
    \begin{subfigure}{0.45\textwidth}
    \raisebox{0.1\height}{\includegraphics[scale=0.8, trim=0 0 0 20]{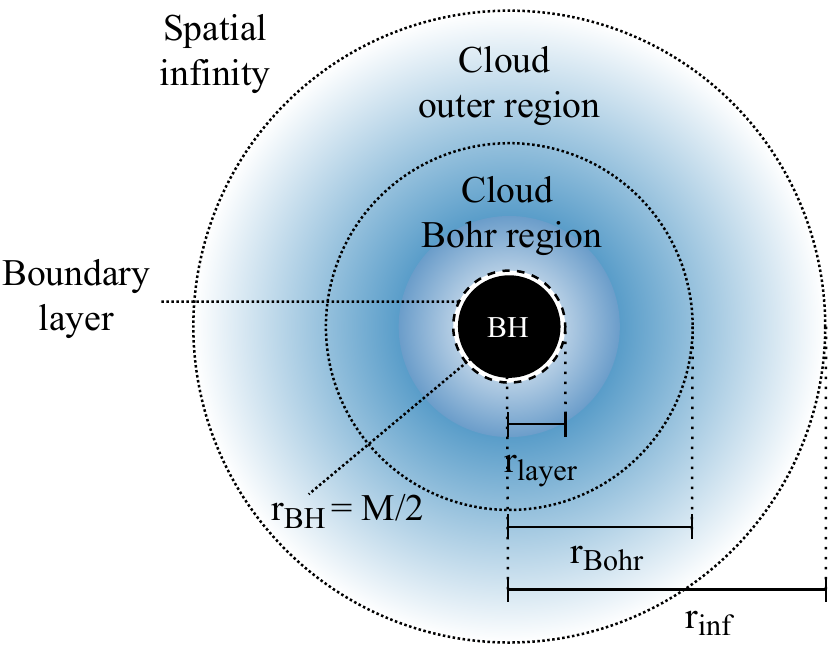}}
    \end{subfigure}
        \caption{\textit{(left)}: Schematic illustration of the dominant eigenstate for the free scalar field, which in the test-field limit has integer labels $\ket{n \ell m} = \ket{211}$, corresponding to a $2p$ configuration. \textit{(right)}: the numerical domain used in this work. The radial boundaries $r_{\rm BH}, r_{\rm layer}, r_{\rm Bohr}$ and $r_{\rm inf}$ separate the black hole horizon, the near-horizon boundary layer, the cloud's Bohr region, the cloud's outer region, and asymptotic infinity (see main text and Table~\ref{table:resolution} for details). For a Schwarzschild background metric, $\tilde{\gamma}_{ij} = \delta_{ij}$ and $N = 0$ at the event horizon is equivalent to choosing the isotropic coordinates, in which case the event horizon is located at $r=r_{\rm BH}=M/2$.
        }
        \label{fig:domain}
\end{figure}

\vskip 4pt

For the scalar field $\Phi$, we demand that it vanishes at large distances. For the two real fields introduced in \eqref{eq:Phitoreal}, this implies
\begin{equation}
        \Phi_1 \vert _{r_{\rm inf}}  = 0 \, , \hskip 56pt \Phi_2 \vert _{r_{\rm inf}} = 0 \, ,  \label{eq:IB_far}
\end{equation}
which leads to an exponentially decaying tail of the boson cloud towards large distances. The boundary condition for the fields at the event horizon, on the other hand, is trickier to impose. To appreciate the subtlety, it is instructive to isolate the radial component of the second derivative term, $\tilde{D}_i \tilde{D}^i \Phi$, in the Klein-Gordon equation (\ref{eq:conformal_KG}). From the boundary condition for the metric quantities at $r=r_{\rm BH}$ given in (\ref{eq:bc_Kadath_paper}), we observe that the coefficient of the radial component of the second-derivative vanishes at the event horizon:
\begin{align}\label{eq:isolated_2nd_OD}
    \Big(\Psi^{-4} - (\beta^r / N)^2\Big) \tilde{D}_r \tilde{D}^r \Phi \hskip 2pt \to \hskip 2pt  0 \, , \qquad \text{as} \qquad r \hskip 2pt  \to \hskip 2pt  r_{\rm BH} \, .
\end{align}
This implies that when approaching the horizon, the radial differential equation dramatically changes from being second order to first order in nature. This behavior of $\Phi$ at the horizon is in fact analytically well understood: dividing the entire Klein-Gordon equation (\ref{eq:conformal_KG}) by this coefficient, the vanishing coefficient (\ref{eq:isolated_2nd_OD}) translates to a divergent pole at the event horizon $r=r_{\rm BH}$ whose residue describes the boundary condition of the event horizon~\cite{Detweiler1980, Baumann2019}. Nevertheless, from a numerical point of view, this so-called degenerate behavior of the differential equation presents important practical challenges, as a naive implemenation of the boundary condition at $r=r_{\rm BH}$ would yield divergent numeric results.

\vskip 4pt

To avoid the numerical instability we introduce a boundary layer, $r=r_{\rm layer}$, at a small distance away from the horizon, on which we impose the inner boundary condition of the scalar field (see Fig.~\ref{fig:domain}). As long as this layer is well within the Compton wavelength of the boson field, $r_{\rm BH} < r_{\rm layer} \ll \mu^{-1}$, imposing the boundary condition at $r_{\rm BH}$ and $r_{\rm layer}$ merely differs by a choice of the amplitude. Motivated by the fact that the complex scalar field respects axisymmetry, we make the ansatz $\Phi \propto e^{i m \phi}$ at the inner boundary, where $\phi$ is the azimuthal coordinate,\footnote{More generally, the complex scalar obeys both axisymmetry and stationarity, $\Phi \propto e^{-i \omega t + i m \phi}$. Without loss of generality, we take $t=0$ for convenience. \label{footnote:ignoret} } and focus on $m=1$ corresponding to the dominant superradiant mode that first saturates the superradiance inequality (\ref{eqn:inequality}). This leads us to impose the following boundary conditions for the two real fields $\Phi_1, \Phi_2$:
\begin{equation}
\begin{aligned}
\label{eq:IB}
    \Phi_1 \vert _{r_{\rm layer}} & = A \hskip 1pt  \cos\phi  \, , \qquad \Phi_2\vert _{r_{\rm layer}} = A \hskip 1pt \sin\phi \, ,
\end{aligned}
\end{equation}
where $A$ is a constant which dictates the mass of the boson cloud. In principle, one could include a polar dependence $\theta$ in the boundary conditions (\ref{eq:IB}). However, in practice, we find that the $\theta$ dependence does not affect our numerical results at all. This is not surprising since spherical symmetry is not an exact isometry of the fully nonlinear system, thus the orbital quantum number $\ell$, which determines the $\theta$-dependence in a spherical harmonic, is not an exact eigenvalue.

\begin{table}[t!]
\centering
\begin{tabular}{c|ccccc}
Region & \begin{tabular}[c]{@{}c@{}} Black\\ hole \end{tabular} & \begin{tabular}[c]{@{}c@{}}Boundary\\  layer \end{tabular} & \begin{tabular}[c]{@{}c@{}}Cloud \\ Bohr region \end{tabular} & \begin{tabular}[c]{@{}c@{}}Cloud \\ outer region \end{tabular} & \begin{tabular}[c]{@{}c@{}}Spatial \\  infinity \end{tabular} \\[10pt] \hline
Outer radius $[M/2]$ & $r_{\rm BH } = 1$ & $r_{\rm layer} = 2.2$ & $r_{\rm Bohr} = 200$ & $r_{\rm inf} = 450$ & - \\
$r$ resolution & -                                                    & 9                                                         & 65                                                         & 55                                                            & 13                                                          \\
$\theta$ resolution   & -                                                    & 17                                                        & 17                                                         & 17                                                            & 17                                                          \\
$\phi$ resolution & -                                                    & 12                                                        & 12                                                         & 12                                                            & 12                                                         
\end{tabular}
\caption{The sizes and the resolutions, i.e. the number of uniformly-spaced grid points, in the radial, polar and azimuthal directions for each region of our numerical domain, c.f. Fig.~\ref{fig:domain}. Radial distances are measured in units of $r_{\rm BH} = M/2$. The cloud outer region $r_{\rm Bohr} < r < r_{\rm inf}$, which emcompasses the exponentially decaying tail of the cloud, is introduced in order to adapt this region to a lower $r$-resolution compared to the cloud Bohr region, where the Bohr radius of the cloud is concentrated. We present convergence studies which led to our choices of resolution above in \S\ref{sec:redshiftpotential_Schw}.}
\label{table:resolution}
\end{table}

\subsubsection{Schwarzschild Approximation} \label{sec:Schwarzschild}

Although superradiance is triggered by a highly-spinning black hole, the end state of this instability process is a slowly-rotating black hole at the center of the boson cloud. The final black hole spin can be inferred by the saturation of the superradiance inequality (\ref{eqn:inequality}), which gives
\begin{equation}
    \tilde{a} \equiv \frac{a}{M} = \frac{4 m (M \omega)}{m^2 + 4 (M \omega)^2} \, .
\end{equation}
where $M \omega \approx M \mu \equiv \alpha$ in the low-frequency limit, see (\ref{eqn:eigenfreq}) below. For the dominant $m=1$ mode, $\tilde{a} \lesssim 0.4$ for $\alpha \lesssim 0.1$. Although the remnant black hole spin would affect the cloud, e.g. introducing mild departures from spherical symmetry, those effects are marginal compared to those introduced by the black hole mass, which provides the dominant source of gravitational potential between the cloud and the central black hole~\cite{Dolan:2007mj, Baumann2019a, Baumann2019}.

\vskip 4pt

As one of the main aims of this work is to investigate the effects of the scalar non-linearities on the gravitational atoms, we shall take $\tilde{a}=0$ in the rest of this paper, such that any visible departures from the free-field cases can be clearly attributed to finite $\lambda$ but not $\tilde{a}$. Taking the Schwarzschild limit also simplifies our numerical computations: it is well known that the Kerr spacetime cannot be embedded in a conformally-flat spatial hypersurface~\cite{Garat2000, Kroon2004}, $\tilde{\gamma}_{ij} = \delta_{ij}$, which as mentioned in \ref{sec:XCTS} is our choice of the initial data input. Without this simplification, we would have to solve for the $\tilde{\gamma}_{ij}$ for the Kerr geometry, which would significantly increase the complexity of our computation.

\vskip 4pt

 To restrict ourselves to a Schwarzschild black hole at the center of the gravitational atom, we use the following simplified boundary conditions in (\ref{eq:bc_Kadath_paper}) for the metric quantities:
\begin{equation}
    \begin{aligned}
        \left( \tilde{D}_r\Psi + \frac{\Psi}{2r} \right) \Big\rvert_{r_{\rm BH}}  = 0 \, , \qquad
        N \rvert_{r_{\rm BH}} = n_0 = 0 \, , \qquad \beta^i \rvert_{r_{\rm BH}} = 0 \, . \label{eq:vacuum_schwarzschild_IB}
    \end{aligned}
\end{equation}
Since $n_0$ is a constant that merely determines the choice of coordinates, we shall take $n_0 = 0$ for convenience. This choice has the added advantage of significantly simplifying the boundary condition for $\beta^i$ at the horizon. In particular, because $\beta^i$ vanishes both at the horizon and at large distances, c.f. (\ref{eqn:bc_metric_inf}), we have the trivial solution $\beta^i=0$. For practical implementations, as long as the correct boundary conditions (\ref{eqn:bc_metric_inf}) and (\ref{eq:vacuum_schwarzschild_IB}) are consistently imposed, we can set $\beta^i=0$ throughout the system of XCTS equations and focus only on the solutions for $\Psi, N $ and $\Phi$. 

\vskip 4pt

For the scalar fields, we impose the boundary conditions as discussed in \S\ref{sec:bc}. With the specializations in (\ref{eq:vacuum_schwarzschild_IB}) the azimuthal symmetry of the initial spacetime is enhanced to spherical symmetry, and the orbital quantum number $\ell$, which characterizes the $\theta$-angular dependence of the field profile, is in principle a definite eigenvalue. However, we find that it remains unnecessary to include a dependence on $\theta$ in the boundary condition (\ref{eq:IB}). This is a consequence of the fact that the $\ell$ dependence of the scalar field eigenstate is already included in the input value of the eigenfrequency (discussed in ~\ref{sec:stationaryfield} below), with the initial guess for the $\theta$-angular profile in (\ref{eq:IB}) having no impact on our numerical solutions.

\subsubsection{Stationarity and Eigenfrequencies} \label{sec:stationaryfield}

In this work, we focus on solutions of boson clouds at the end state of the superradiant amplification process, i.e. the superradiance inequality (\ref{eqn:inequality}) is saturated and the cloud stops growing. This allows us to restrict ourselves to stationary boson field configurations, which satisfy
\begin{equation}
    \partial_t \Phi = - i \omega \Phi \, . \label{eqn:Lietime}
\end{equation}
where $\omega$ is the eigenfrequency. Since the imaginary part of the eigenfrequency vanishes when the superradiance inequality is saturated~\cite{Detweiler1980, Dolan:2007mj, Baumann2019}, $\omega$ is a real quantity.

\vskip 4pt

The stationarity condition (\ref{eqn:Lietime}) recasts the problem of solving the Klein-Gordon equation (\ref{eq:conformal_KG}) into an eigenvalue problem. As with all eigenvalue problems, there is in principle an infinite tower of eigenstates which satisfy the boundary conditions (\ref{eq:IB_far}) and (\ref{eq:IB}), with each eigenstate corresponding to a different value of $\omega$. In order to study a specific eigenstate in our numerical setup, we would have to input the relevant eigenvalue into the XCTS system of equations. If an exact solution for $\omega$ is not readily available, we should at least choose a $\omega$ that is as close as possible to the eigenstate which we intend to investigate. Empirically, we find that a 'wrong' value for $\omega$ either yields a solution of $\Phi$ that is far away from the desired state or generates divergent numerical results. This sensitivity to $\omega$ implies that accurate eigenvalues are critical to obtaining our desired solutions. 

\vskip 4pt

In the test-field limit, where the backreaction of the boson cloud on the metric is negligible, the eigenfrequencies for the free scalar field ($\lambda=0$) can be solved numerically for arbitrary values of $\mu$~\cite{Dolan:2007mj, Baumann2019}. However, for a complex scalar field with $\lambda \neq 0$, the eigenfrequencies are to date unknown because the eigenvalue problem in this case is non-linear, which is technically more challenging to solve (see~\cite{Baryakhtar:2020gao} for analytic calculations of the eigenfrequencies for small $\lambda$ and small $M \omega$ for a real scalar field). In principle, the $\lambda \neq 0$ eigenvalues could be obtained either through a brute-force grid scanning over reasonable values of $\omega$ or solved through standard numerical recipes, e.g. the Newton-Rhapson method. However, we find either of these approaches to be computationally prohibitive, especially because a high resolution for $\omega$ is necessary in order to achieve convergent results. Making no pretense of solving $\omega$ for the $\lambda \neq 0$ scenarios, in this work we choose $\omega$ to be the eigenfrequency of the dominant free scalar field in the test limit, $\ket{n \ell m} = \ket{211}$ for both $\lambda =0$ and $\lambda \neq 0$ cases. Here $n \geq \ell + 1$ is the principal quantum number, $\ell$ is the orbital angular momentum number, and $m$ the azimuthal number as found in (\ref{eqn:inequality}). For $M \omega \lesssim 0.3$, the eigenfrequencies for the $\ket{211}$ free scalar field can be described accurately by the analytic expression~\cite{Baumann2019a, Baumann2019}
\beq
\begin{aligned}
 \omega_{211} (\lambda = 0)  &= \mu \left( 1 - \frac{\alpha^2}{8} - \frac{129 \hskip 1pt \alpha^4}{128} + \frac{2 \hskip 1pt \tilde a  \alpha^5}{3}  + \cdots \right) \, , \label{eqn:eigenfreq}
\end{aligned} 
\eeq
where the dimensionless parameter
\begin{equation}
\alpha \equiv M \mu \, ,
\end{equation}
is the gravitational fine structure constant. For larger values of $\alpha$, we use the numerical solver in~\cite{Baumann2019} to compute $\omega_{211}$.\footnote{In fact, the size of our numerical domain, with the details outlined in Table~\ref{table:resolution}, limits us to boson clouds with moderately large values of $\alpha \gtrsim 0.25$.} Remarkably, as we shall discuss in Section~\ref{sec:numerics}, we find that our numerical results for $\lambda=0$ converge even after backreaction of the cloud on the metric is taken into account. This demonstrates that (\ref{eqn:eigenfreq}) remains accurate for the free scalar field in the strong gravity regime. For $\lambda \neq 0$, our numerical solutions converge with $\omega = \omega_{211}$ for small values of $|\lambda|$. For larger departures from $\lambda=0$, the numerics diverge, indicating that the eigenfrequency (\ref{eqn:eigenfreq}) is no longer accurate for those solutions (see Fig.~\ref{fig:paramspace} for a summary).

\subsubsection{Numerical Evaluation} \label{sec:steps}

The full system of equations described in \S\ref{sec:XCTS}, coupled with the ingredients described from \S\ref{sec:bc} to \S\ref{sec:stationaryfield}, are difficult to solve simultaneously as the numerical eigenvalue solver in \texttt{KADATH}~\cite{Grandclement2010} requires the inversion of a large matrix operator, whose size depends on the size of the domain (right of Fig.~\ref{fig:domain}) and judicious choices for the resolution. To simplify our calculation, we shall focus on the gauge-invariant ADM mass of the backreacted spacetime as a measure of strong gravity effects. We perform our computation through the following procedure:

\begin{enumerate}[leftmargin=15pt]
\item \label{item:1} \textit{Schwarzschild background metric:} Using the boundary conditions described in \S\ref{sec:bc} and \S\ref{sec:Schwarzschild}, we first solve for the metric quantities $\Psi$ and $N$ by ignoring $\Phi$, i.e. restricting ourselves to Einstein's equation in vacuum. Throughout this paper, we will denote the lapse and conformal factor evaluated on this fixed unbackreacted background spacetime as $\overline{N}$ and $\overline{\Psi}$, in order to distinguish them from their fully nonlinear counterparts. 

As described in \S\ref{sec:Schwarzschild}, our boundary conditions for $\beta^i$ would yield the trivial solution $\beta^i=0$, which we tested and confirmed numerically. Since we also impose conformal flatness, $\tilde{\gamma}_{ij} = \delta_{ij}$, the line element of our background metric is equivalent to the Schwarzchild black hole in \textit{isotropic coordinates} $\{t, r, \theta, \phi\}$, which reads
\begin{equation}\label{eqn:iso}
    \d s^2=-\left(\frac{2r-M}{2r+M} \right)^2 \d t^2 + \left(1 + \frac{M}{2r} \right)^4
    \left(\d r^2+r^2 \hskip 1pt \d \theta^2 + r^2 \sin^2 \theta \hskip 1pt \d \phi^2 \right) \, ,
\end{equation}
where $M$ is the black hole mass. The spatial hypersurface of (\ref{eqn:iso}) is clearly isometric to Euclidean space. In addition, unlike the solution expressed in Schwarzschild coordinates, the event horizon in isotropic coordinates is located at $r_{\rm BH}=M/2$. The analytic solutions for $\overline{\Psi}$ and $\overline{N}$ in isotropic coordinates, denoted by $\overline{\Psi}_{\rm ana}$ and $\overline{N}_{\rm ana}$, are also straightforward to derive:
\begin{equation}\label{eq:Schwarzschild_solution_analytic}
    \overline{\Psi}_{\rm ana} = 1 + \frac{M}{2r} \, , \qquad     \overline{N}_{\rm ana} = \frac{2r-M}{2r+M} \, .
\end{equation}
From (\ref{eq:Schwarzschild_solution_analytic}), we see that $\overline{\Psi}$ can be clearly interpreted as the gravitational potential sourced by the black hole, while $\overline{N}$ is the redshift factor. We compare our numeric results with (\ref{eq:Schwarzschild_solution_analytic}) in Section~\ref{sec:numerics} and find excellent agreement.

\item \textit{Boson cloud on the background spacetime}: 
\label{item:2}We solve for the dominant $m=1$ mode using the background metric computed in step \ref{item:1}. In addition to the boundary conditions (\ref{eq:IB_far}) and (\ref{eq:IB}), we use the eigenfrequency of the $\ket{n \ell m} = \ket{211}$  state of the free scalar field in all of the $\lambda = 0$ and $\lambda \neq 0$ computations, as discussed in \S\ref{sec:stationaryfield}. For the $\lambda = 0$ cases, we show in Section~\ref{sec:numerics} that when the backreaction of the cloud on the metric is neglected, our numeric results are in
good agreement with the analytic expression for the $\ket{211}$ eigenfunction
\begin{equation}
    \Phi_{211} (\lambda = 0) \propto \bar{r} \hskip 1pt e^{-\sqrt{\mu^2 - \omega^2} \hskip 1pt \bar{r}} \hskip 1pt e^{i \phi}  \sin{\theta}\, ,  \qquad \bar{r} = r \left( 1 + \frac{M}{2r}  \right)^2 \, , \label{eq:2pR}
\end{equation}
where $\bar{r}$ is the Schwarzschild radial coordinate and we take $t=0$ without loss of generality, c.f. Footnote~\ref{footnote:ignoret}. For ease of comparison between the shapes of scalar field profiles with different amplitudes, we introduce the normalized scalar field
\beq
\Phi^{N} = \Phi / A_r \, , \label{eqn:normalized_field}
\eeq
where $A_r$ is the area under the curve along the radial direction. 

\item \textit{Backreaction of boson cloud on the metric:} \label{item:3} Using the solution for $\Phi$ found in the previous step, we compute the matter variables $\{\rho, j^i, S\}$ through (\ref{eqn:mattervars}) and solve for the nonlinear $\Psi$ and $N$ with the full XCTS equations. These backreacted metric variables show departures from the background solution~(\ref{eq:Schwarzschild_solution_analytic}) and contain information about the gravitational atom. For instance, the ADM mass~\cite{Gourgoulhon:2007ue}
\begin{equation}\label{eq:ADM_mass}
    M_{\text{ADM}} = -\frac{1}{2\pi}\oint_{r =r_{\rm inf}} D_i\Psi \, \text{d}S^i\, , 
\end{equation}
a gauge-invariant quantity evaluated at the surface $S^i$ at $r = r_{\rm inf}$, includes the total energy of the black hole mass \emph{and} the energy of the boson cloud. Since $M$ is the black hole mass in the absence of the cloud, we define the mass of the boson cloud as
\begin{equation}
    M_c = M_{\rm ADM} - M \, , \label{eqn:cloud_mass}
\end{equation}
which serves as a useful parameter to quantify the degree to which backreaction is important.

\end{enumerate}
\noindent The solutions for each of these steps will be shown in Section~\ref{sec:numerics}.

\pagebreak

\section{Numerical Results} \label{sec:numerics}

In this section, we present numerical solutions to the setup described in Section~\ref{sec:setup}. We first discuss the results obtained for the Schwarzschild background metric and the free scalar field in \S\ref{sec:SchwBackground} since analytical results are known in these cases. This serves as a testbed to check of the outputs from the code against the analytics and further enables us to quantify the regime of validity of the analytical results. Then, in \S\ref{sec:backreaction}, we consider results for scenarios that include the gravitational backreaction of the boson cloud onto the background spacetime and  self-interactions of the scalar field, for which no analytical results are known in the literature. 

\subsection{Testing Numerics against Analytics} \label{sec:SchwBackground}

Here we present for the Schwarzschild metric quantities, lapse $\overline{N}$ and conformal factor $\overline{\Psi}$, and the free field scalar field $\Phi$ on the Schwarzschild metric. Since analytical expressions are known for these quantities (at least for $\alpha \ll 1$ for $\Phi$), we can compare our numerical outputs with the analytical results. As we shall see, these results will also form the foundation for the self-interacting scalar field and fully nonlinear backreaction results in \S\ref{sec:backreaction}.

\subsubsection{Schwarzschild Spacetime} \label{sec:redshiftpotential_Schw}

\begin{figure}[h!]
\centering
\includegraphics[width=\textwidth]{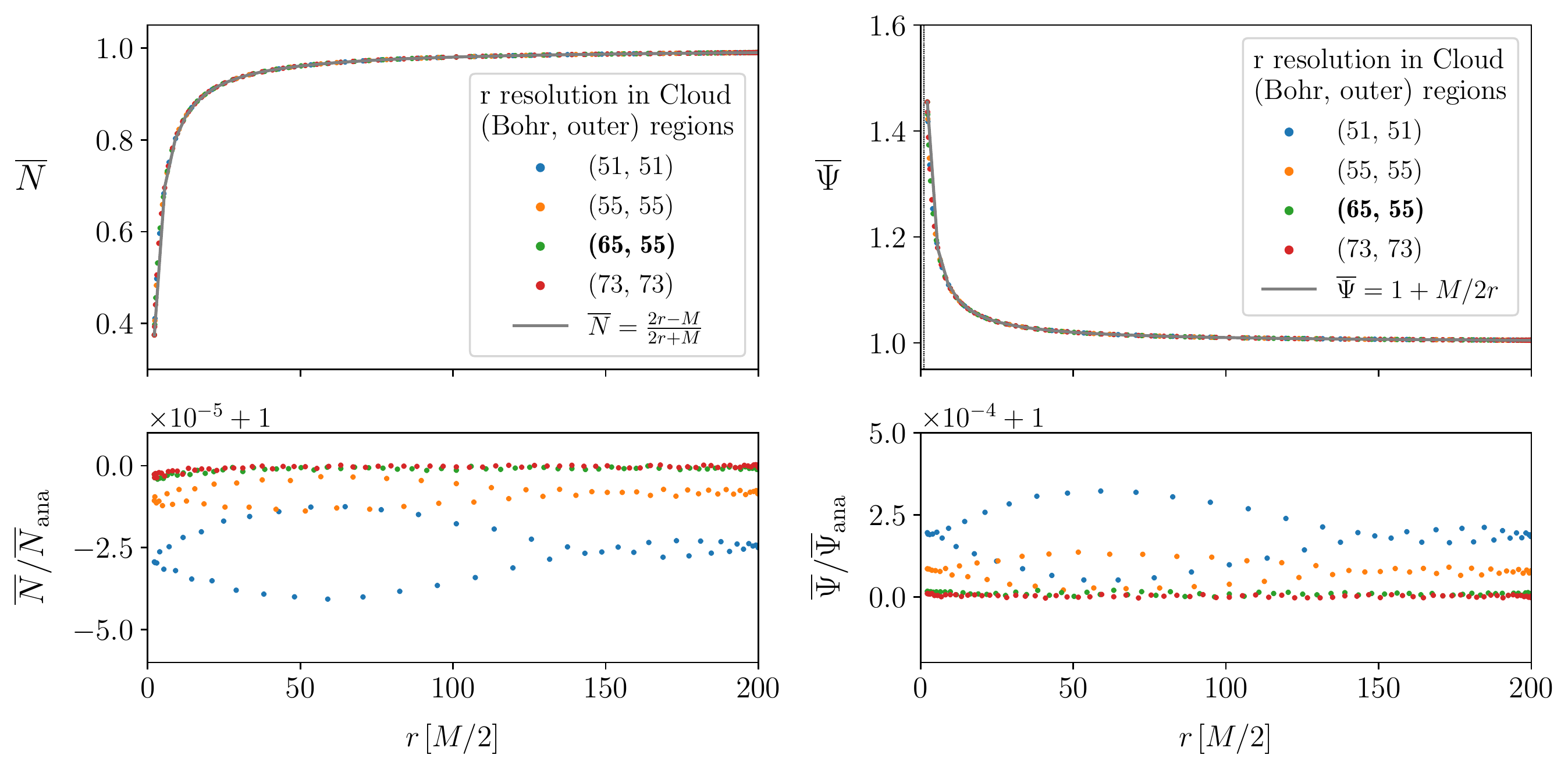}
    \caption{The lapse function $\overline{N}$ \textit{(left)} and conformal factor $\overline{\Psi}$ \textit{(right)} of the Schwarzschild background metric for a range of radial resolution choices in the cloud (Bohr, outer) regions; cf. Fig.~\ref{fig:domain}. Striking a balance between computational efficiency and numerical precision, we use the $r$ resolution of $(65, 55)$ throughout this work (\textit{bold}), cf. Table~\ref{table:resolution}. The bottom panels show the ratio of those numeric data with the analytic expression (\ref{eq:Schwarzschild_solution_analytic}), demonstrating the convergence of the numerics to the analytics.}
    \label{fig:conflapse}
\end{figure}

Figure~\ref{fig:conflapse} shows the numerical results for $\overline{N}$ and $\overline{\Psi}$ as a function of $r$ on the spatial slice $\phi=0$ for different choices of resolutions in the two domains illustrated in Fig.~\ref{fig:domain}. We first notice that the general behavior of the functions is as expected. The lapse, corresponding to the redshift factor, is small near the horizon, indicating that coordinate time passes slowly, and increases to its Newtonian value $\overline{N} \to 1$ as $r \to \infty$. Note that the plot focuses on the $r < r_{\rm Bohr}$ region, cf. Fig.~\ref{fig:domain}, at which point the metric functions are already close to their asymptotic value. Outside of the region shown in the plot, $ \overline{N} \to 1$ limit would continue at a slower rate. Likewise, near the black hole horizon the conformal factor is very different from its asymptotic value $\overline{\Psi} \to 1$. The most prominent differences from the asymptotic values occur at distances smaller than $\sim 40$ Schwarzschild radii from the black hole. At larger distances, a small but noticeable discrepancy remains, which decays away very slowly towards larger radii. 

\vskip 4pt

With regards to numerical resolution, we find that the numerics converge as the domain resolutions increase and asymptotically approach the analytic expressions $\overline{N}_{\rm ana}$ and $\overline{\Psi}_{\rm ana}$ in (\ref{eq:Schwarzschild_solution_analytic}), as expected from the single-domain setup~\cite{Grandclement2010}. Below each of the subplots in Fig.~\ref{fig:conflapse}, we plot the ratio of the numerics and the analytics. We find that $65$ and $55$ number of resolution points in the cloud Bohr and outer regions, respectively, robustly achieve $\sim 10^{-3} - 10^{-4}$ level accuracy for these metric quantities. This result forms the basis for our choice of resolution, as shown in Table~\ref{table:resolution}. As an additional check, we computed the ADM mass  (\ref{eq:ADM_mass}) with these numerical data, finding that $M_{\text{ADM}}= M$ as expected for the Schwarzschild black hole, up to the same level of precision. 

\subsubsection{Free Scalar Cloud Profiles}

Using the numerical resolution deduced in \S\ref{sec:redshiftpotential_Schw} for the metric quantities, we now compare the numerical and analytical results for the free scalar field. Since analytic expressions for a self-interacting scalar field profile are not known in the literature, we will omit the interactions from this comparison study.


\begin{figure}[b!]
\centering
	\begin{subfigure}{.48\textwidth}
  	\includegraphics[width=\textwidth, trim=10 0 0 0]{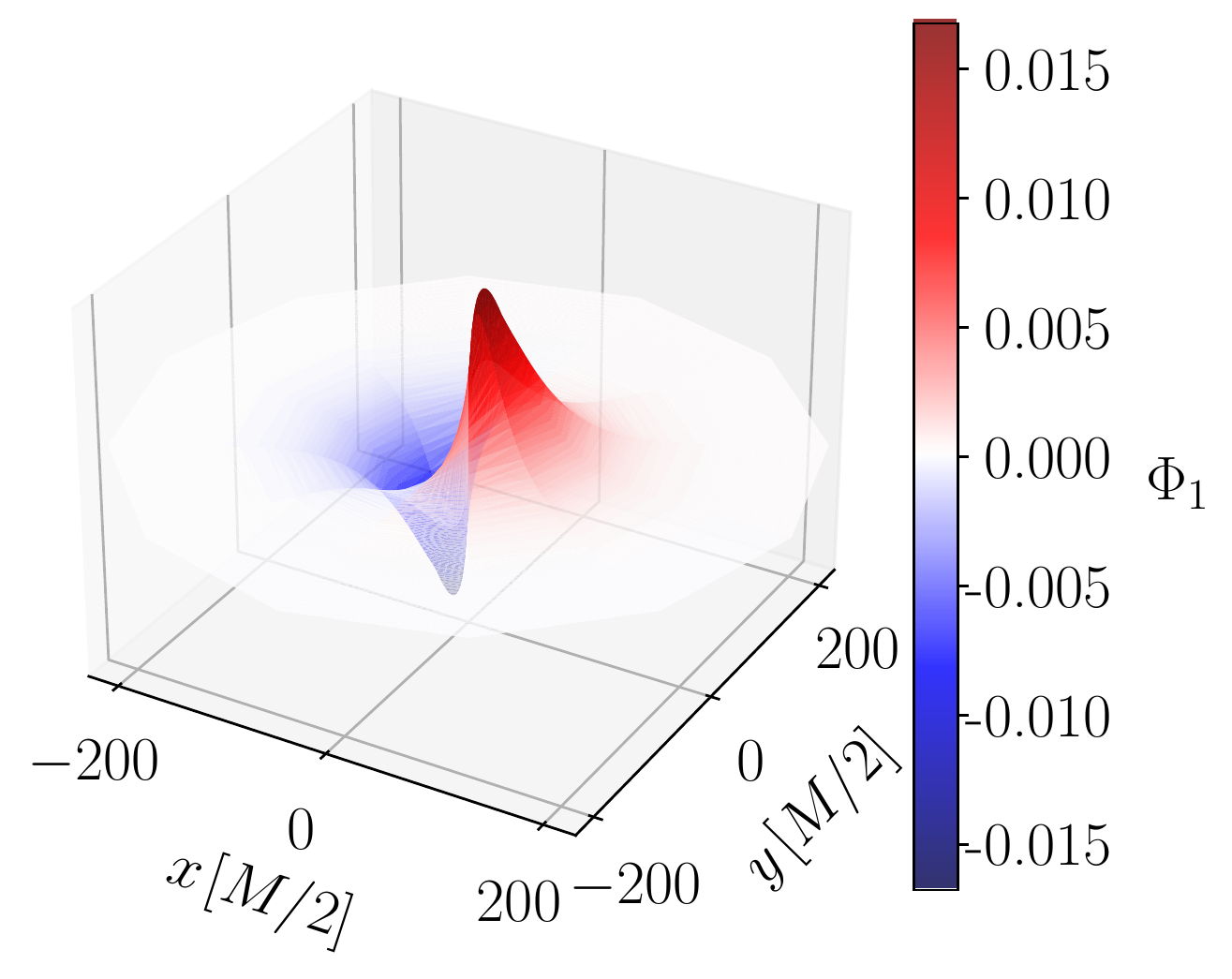}
	\end{subfigure}
	\begin{subfigure}{.48\textwidth}
 	 \includegraphics[width=\textwidth, trim=10 0 0 0]{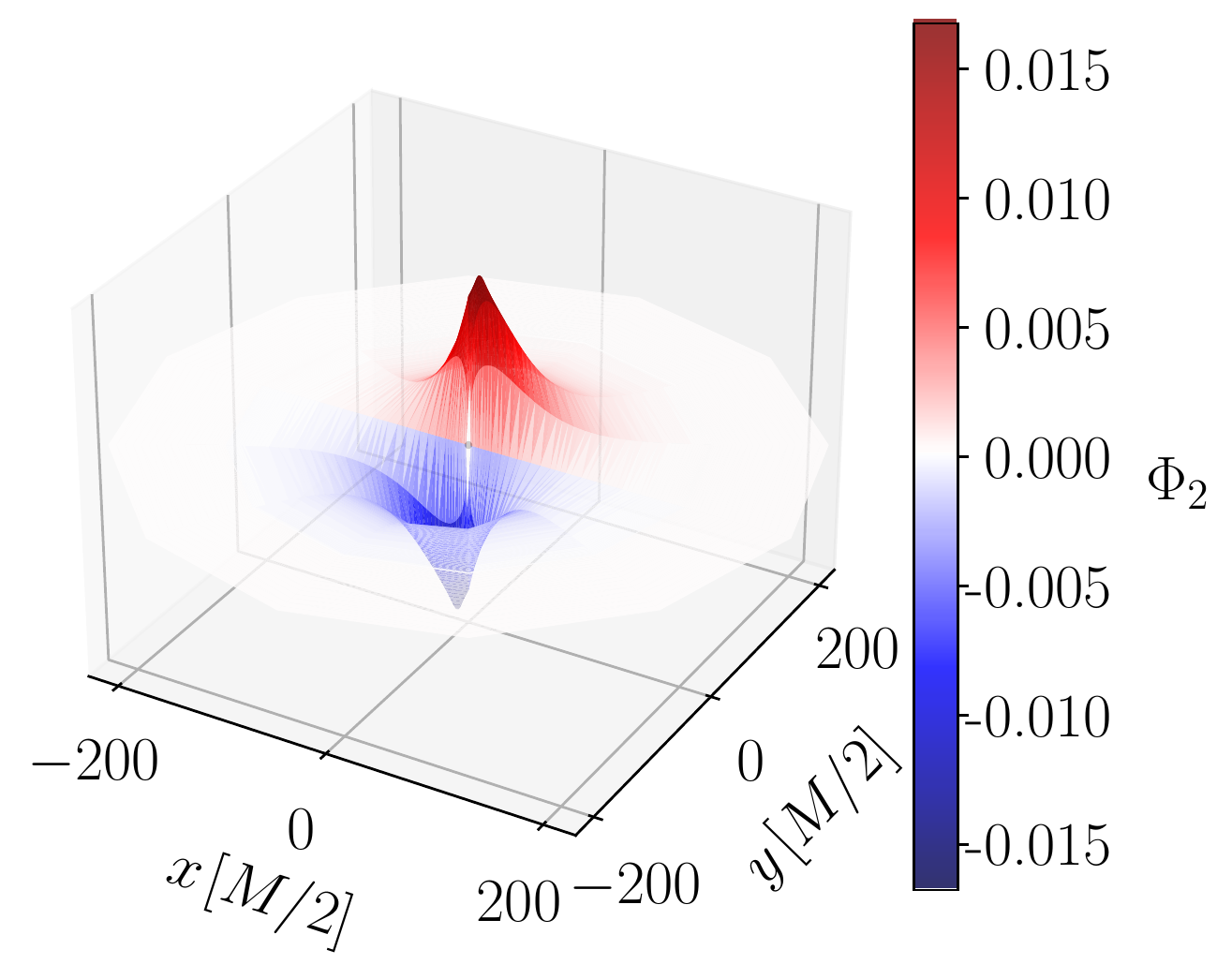}
	\end{subfigure}
	\caption{The free scalar field configuration $\Phi_1$ \textit{(left)} and $\Phi_2$ \textit{(right)} on a Schwarzschild background metric, projected onto the two-dimensional equatorial plane, $\theta=\pi/2$.  $\Phi_1$ and $\Phi_2$, which are the real and imaginary parts of the complex field $\Phi$, cf. (\ref{eq:Phitoreal}), differ by a $\pi/2$ rotation around the $z-$axis. These scalar fields are normalized by dividing them with the interpolated area under the field profiles.}
	\label{fig:3d_plots}
\end{figure}

\vskip 4pt

In Fig.~\ref{fig:3d_plots}, we show the results for the free-field scalar clouds without backreaction onto the Schwarzschild metric. On the left panel, we present the two-dimensional projection of $\Phi_1$ on the equatorial plane, i.e. $\theta=\pi/2$; see (\ref{eq:Phitoreal}) to recap the definition of $\Phi_1$ as the real part of the complex field $\Phi$. The solution for $\Phi_2$, which is the imaginary part of $\Phi$, differs from $\Phi_1$ through a $\pi/2$ rotation around the $z-$axis and is shown on the right panel. These field profiles are normalized appropriately by dividing the numeric data with the interpolated area under the curve, see (\ref{eqn:normalized_field}). 

\begin{figure}[t!]
    \centering
     \includegraphics[width=.6\textwidth]{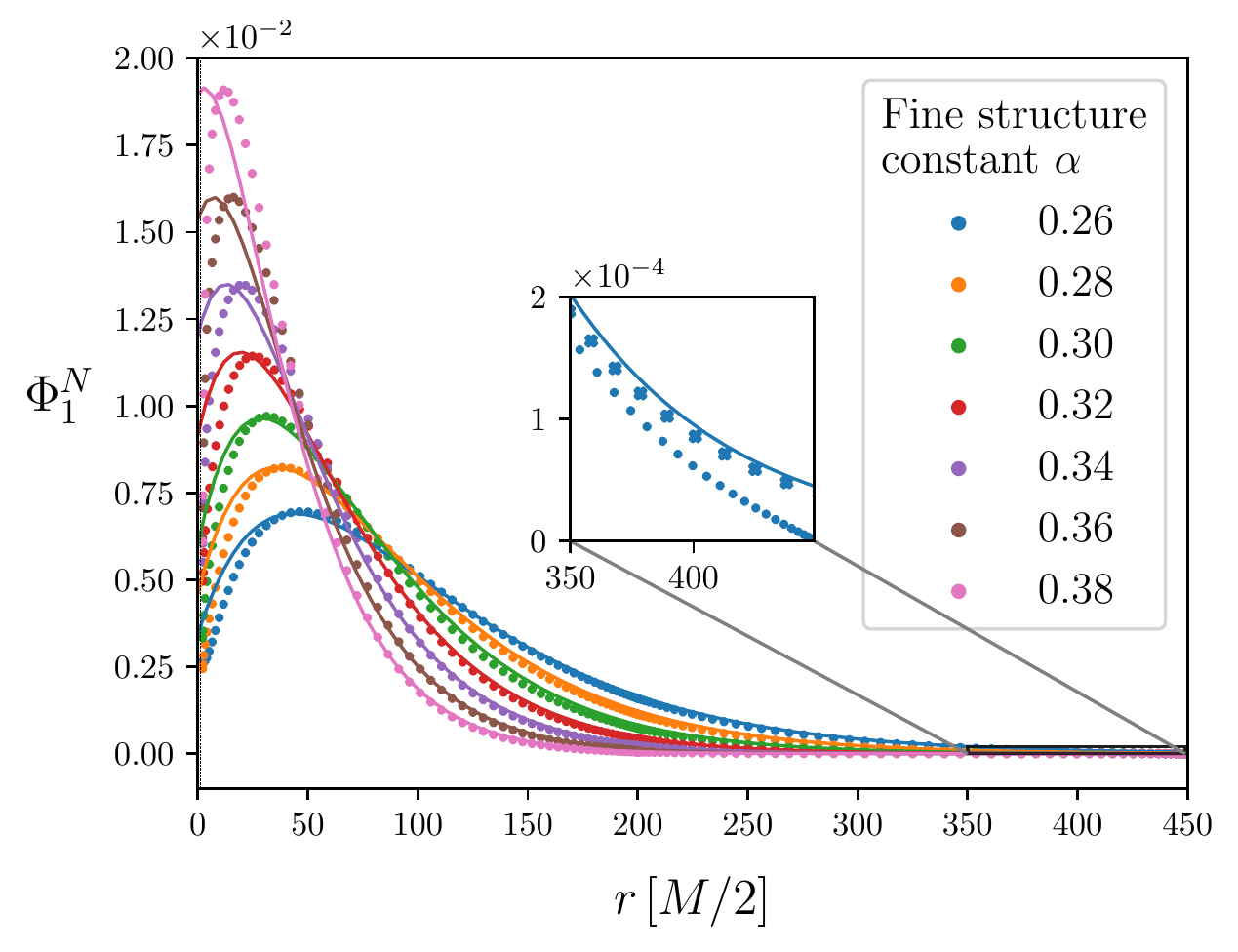}
    \caption{Numerical results (dotted) of the free scalar field for different values of $\alpha$ compared with the analytical expression (\ref{eq:2pR}) (solid), which is strictly valid only for $\alpha \ll 1$. The deviation between analytics and numerics near the horizon, especially for large values of $\alpha$, arise due to nonlinear gravitational interactions between the boson cloud and the black hole that is not captured by the analytics. In the inset, we focus on the large distance behaviour of the $\alpha=0.26$ profile, finding that the analytics agree with the numerics computed with an outer boundary size of $r_{\rm inf}=800 \hskip 1pt [M/2]$ (crosses), while our default setup with $r_{\rm inf}=450 \hskip 1pt [M/2]$ (dotted) deviates from the peak value by $\sim 1\%$, a small effect that does not significantly affect the boundary condition at $r_{\rm inf}$. See main text for more discussion.}
    \label{fig:different_mu}
\end{figure}

\vskip 4pt

\begin{figure}[t!]
    \centering
     \includegraphics[width=\textwidth]{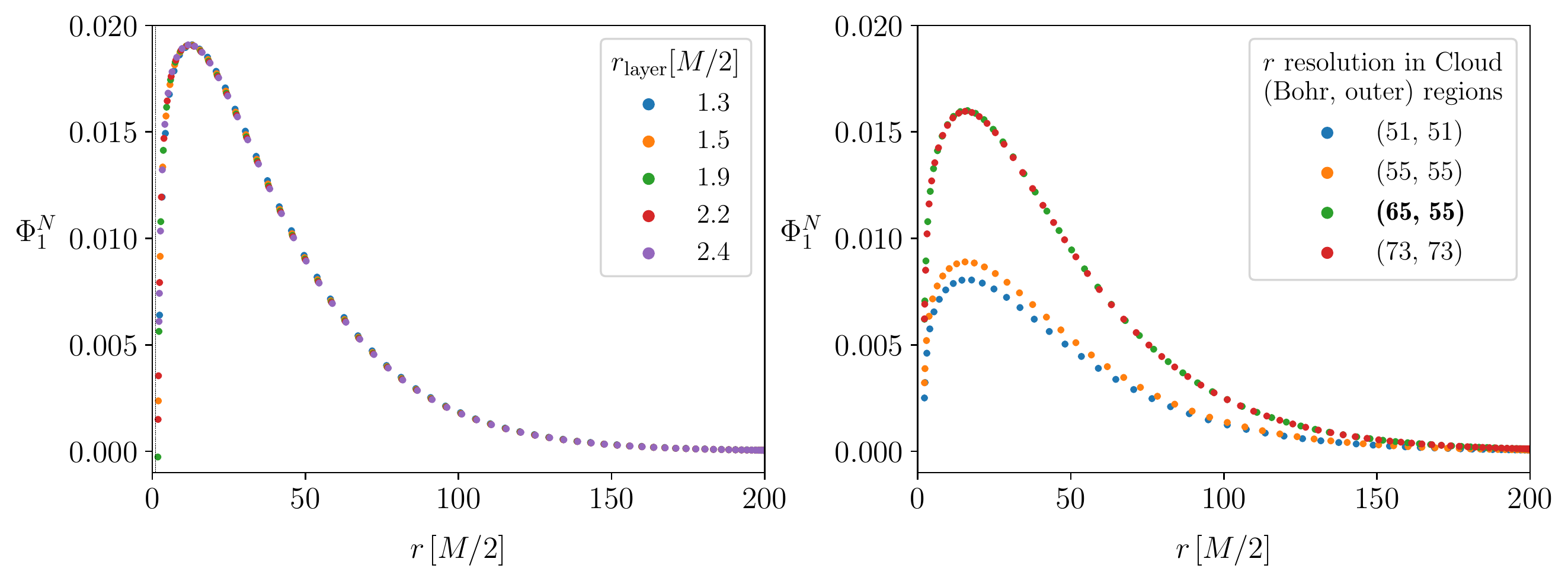}
    \caption{(\textit{left}) 
    A systematic study for the effects of the boundary layer size, $r_{\rm layer}$, on the scalar field profile. In this figure we set $\alpha=0.38$, $\lambda=0$ and plot the normalized scalar field (\ref{eqn:normalized_field}) for easy comparison of the shapes of the curves. We see that resulting field configurations for different values of $r_{\rm layer}$ coincide. As argued in the main text, this is expected as long as $r_{\rm BH} < r_{\rm layer} \ll \mu^{-1}$, which in this example corresponds to $1 < r_{\rm layer} [M/2] \ll 5.3$. Some of these field profiles exhibit zero crossing near the horizon, which are simply artefacts introduced by $r_{\rm layer}$. (\textit{right}) Convergence study of the scalar field profile for $\alpha=0.36$ and $\lambda=0$ with increasing $r$ resolution in the Cloud (Bohr, outer) regions, cf. Fig.~\ref{fig:conflapse} for the analog for $\overline{\Psi}$ and $\overline{N}$.}
    \label{fig:r1}
\end{figure}

Figure~\ref{fig:different_mu} shows the projection of $\Phi_1$ onto the equatorial plane and along the $x-$axis, i.e. $\theta = \pi/2 $ and $\phi = 0$. For a comprehensive comparison with the analytic expression (\ref{eq:2pR}), we demonstrate the numeric solutions for a range of values of $\alpha \in [0.26, 0.28, 0.30, 0.32, 0.34, 0.36, 0.38]$. The lower bound on $\alpha$ is constrained by the size of our numerical domain, while the upper bound is close to the $\alpha \approx 0.4$ limit, beyond which superradiance of a spinning black hole would not grow the cloud~\cite{Detweiler1980, Baumann2019}. To compare with the analytics, we additionally plot the analytic expression (\ref{eq:2pR}), which is strictly valid only for $\alpha \ll 1$.\footnote{Since (\ref{eq:2pR}) is strictly valid for $\alpha \ll 1$ in the ``far region", defined as the region between $ M  \lesssim r < \infty$~\cite{Baumann2019}, we manually fit the analytic curves with the numerics by matching their values at the exponential tail at large distances, approximately at $r \gtrsim 70 (M/2)$ in Fig.~\ref{fig:different_mu}.  
} 
A few trends clearly emerge from these comparisons:

\begin{enumerate}

\item The analytic expression (\ref{eq:2pR}), which is valid for $\alpha \ll 1$, does not exactly agree with the numerics even for our smallest value of $\alpha = 0.26$, especially in the region between the Bohr peak and the event horizon. This indicates that strong-gravity effects are already present in the near-horizon region for $\alpha$ as small as $0.26$. 

While the discrepancy between analytics and numerics could arise due to the introduction of the boundary layer $r_{\rm layer}$ in our numeric domain, cf. Fig.~\ref{fig:domain}, we rule out this hypothesis in based on the findings when varying $r_{\rm layer}$ presented in the left panel of Fig.~\ref{fig:r1}, which shows that different values of $r_{\rm layer}$ have no impact on the solution. The insensitivity to $r_{\rm layer}$ is expected because, as we argued around (\ref{eq:IB}), as long as $r_{\rm BH} < r_{\rm layer} \ll \mu^{-1}$ different choices of $r_{\rm layer}$ simply corresponds to different choices of $A$ -- a constant that will be removed through our appropriate normalization. 

\item As $\alpha$ increases, the Bohr region is peaked closer to the horizon, and the deviation between the analytics (\ref{eq:2pR}) and the numerics in Fig.~\ref{fig:different_mu} becomes more prominent. This arises because as $\alpha$ increases the nonlinear gravitational coupling between the boson field and black hole and pushes the Bohr peak closer to the event horizon.
    
\end{enumerate}

\noindent Finally, as a consistency check, we show the results of the convergence study for $\Phi_1$ in the inset of Fig.~\ref{fig:different_mu} and the right panel of Fig~\ref{fig:r1}. In the former, we show the impact of the spatial infinity domain size, $r_{\rm inf}$, on the fall-off behaviour of the cloud to achieve the boundary condition (\ref{eq:IB_far}) at large distances. There we find that even for the smallest $\alpha=0.26$ studied, whose Bohr radius is the largest, which makes it most prone to errors at $r_{\rm inf}$, achieves a relative accuracy of $\sim 10^{-3}$ compared to the peak value of $\Phi_1$. For larger values of $\alpha$, the error at $r_{\rm inf}$ decreases rapidly by orders of magnitude. We thus conclude that the boundary condition that we impose at large distances is sufficiently accurate for our purposes. On the other hand, similar to $\overline{N}$ and $\overline{\Psi}$ in Fig.~\ref{fig:conflapse}, we find that the numerical solutions converge when the number of radial grid points in the cloud (Bohr, outer) regions are above $(65, 55)$, respectively. In this work, we balance numerical accuracy and computational efficiency by adopting the resolution $(65, 55)$.

\subsection{Effects of Scalar Self-Interactions} \label{sec:backreaction}

With the robust understanding of our numerical results developed in the previous section, we will now present the self-interacting scalar field profile in \S\ref{sec:interacing_scalars} and the effects of backreaction of the cloud onto the spacetime metric in \S\ref{sec:backreaction_solution}. In \S\ref{sec:paramspace}, we show a comprehensive sampling of the $M_c - \lambda$ parameter space in order to understand our solutions and the regime of validity of our numeric methods. There we also discuss the scalings of $M_c$ with $A$ and $\lambda$ in our numerics.

\subsubsection{Self-Interacting Scalar Cloud Profiles} \label{sec:interacing_scalars}

\begin{figure}[b!]
    \centering
    \includegraphics[width=\textwidth]{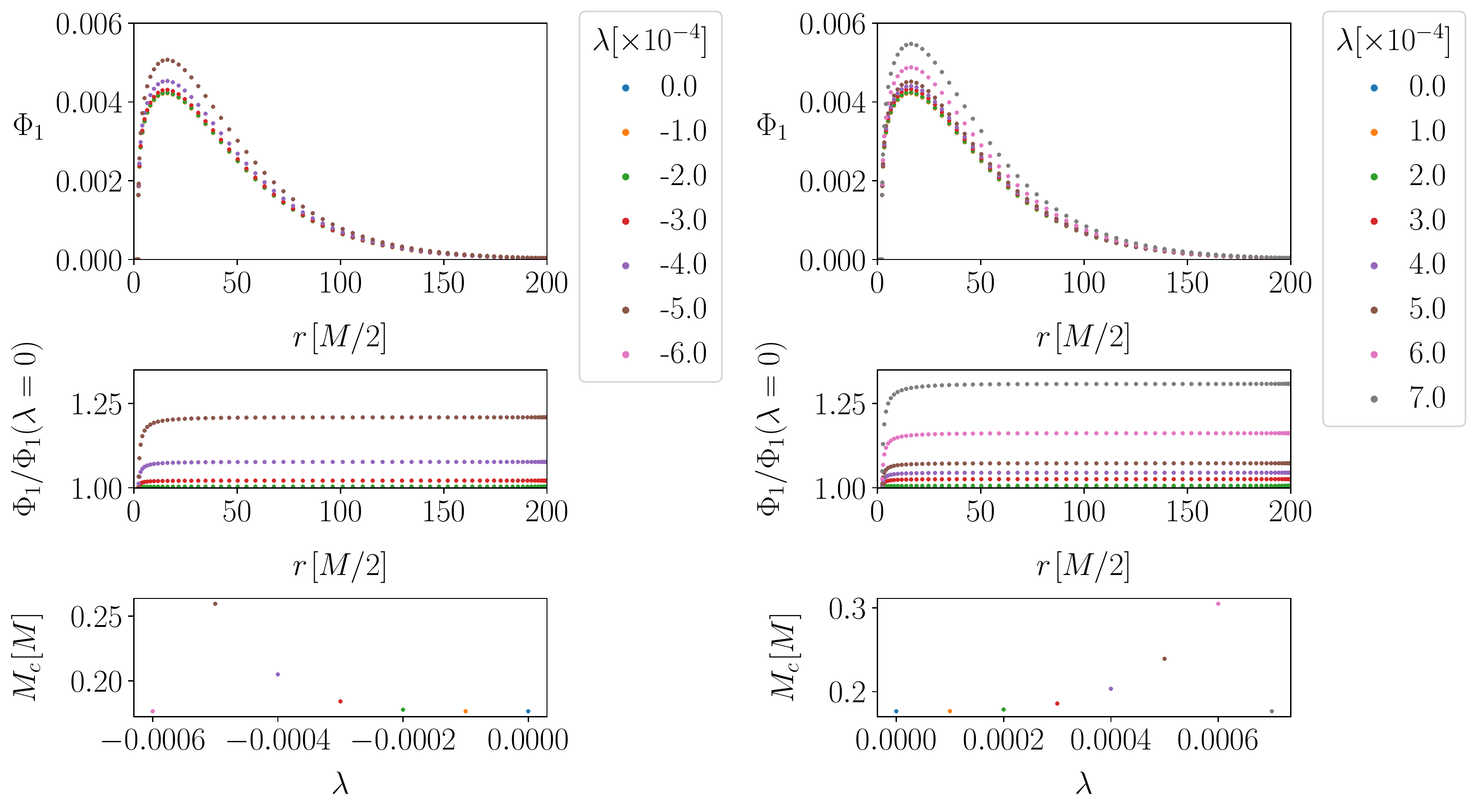}
    \caption{
    \textit{(left)}: The field profile $\Phi_1$ for an attractive $\lambda < 0$ self-interacting potential with $A=0.0006$ and $\alpha=0.36$. Here we show $\Phi_1$, instead of the normalized field $\Phi_1^N$, in order to illustrate the effect of self interactions on the total scalar field amplitude. \textit{(right)}: Same as the left panel except we show the profiles for a repulsive $\lambda > 0$ potential. The middle panels show the ratio between the self-interacting scalar profiles with their $\lambda = 0$ free-field counterpart. 
    The lowest panel demonstrates the mass of the cloud, $M_c$, as a function of $\lambda$. In the bottom panels we also include points beyond  $ -5 \times 10^{-4} \lesssim \lambda \lesssim 6 \times 10^{-4}$, indicating that the $M_c - \lambda$ curve is no longer monotonic and that our approximations in \S\ref{sec:bc} are no longer reliable beyond this range for this choice of $A$.}
    \label{fig:phi_lambda}
\end{figure}

The self-interacting scalar field configurations for a range of values for $\lambda$ are shown in Fig.~\ref{fig:phi_lambda}, where the left panel of Fig.~\ref{fig:phi_lambda} illustrates the solutions for the $\lambda < 0$ attractive potential while
 the right panel focuses on the $\lambda > 0$ repulsive cases. For reference we also include the $\lambda=0$ curves discussed in the previous subsection. In both scenarios, we see that the total field amplitude increases as the magnitude of $\lambda$ increases, as highlighted in the middle panels of Fig.~\ref{fig:phi_lambda} through the ratios $\Phi(\lambda \neq 0) / \Phi(\lambda = 0)$. This is further illustrated in the bottom panels where we observe $M_c$ scales approximately quadratically with $\lambda$. The fact that $M_c$ scales approximately quadratically, instead of linearly, with $\lambda$ suggests that second order effects in $\Phi$ have started to become important for $|\lambda| \sim 10^{-4}$ (see \S\ref{sec:paramspace} below for more discussions on the scalings with $\lambda$ and $A$). This scaling can lead to substantial differences between the mass of a free scalar cloud and that of a self-interacting scalar cloud. For example, we find $M_c = 0.304M$ for $\lambda=0.0006, \alpha=0.36, A = 0.0006$, which is a $72\%$ increase from the $\lambda=0$ cloud mass with the same $\alpha$ and $A$. For $ -5 \times 10^{-4} \lesssim \lambda \lesssim 6 \times 10^{-4}$ in both cases the monotonic increase between $M_c$ and $\lambda$ is broken, suggesting that our approximations in \S\ref{sec:bc} are no longer reliable beyond this range. A comprehensive exploration of solutions over the $M_c-\lambda$ parameter space will be given in \S\ref{sec:paramspace}.

\subsubsection{Spacetime with Cloud Backreaction} \label{sec:backreaction_solution}

\begin{figure}[b!]
\centering
\includegraphics[width=\textwidth]{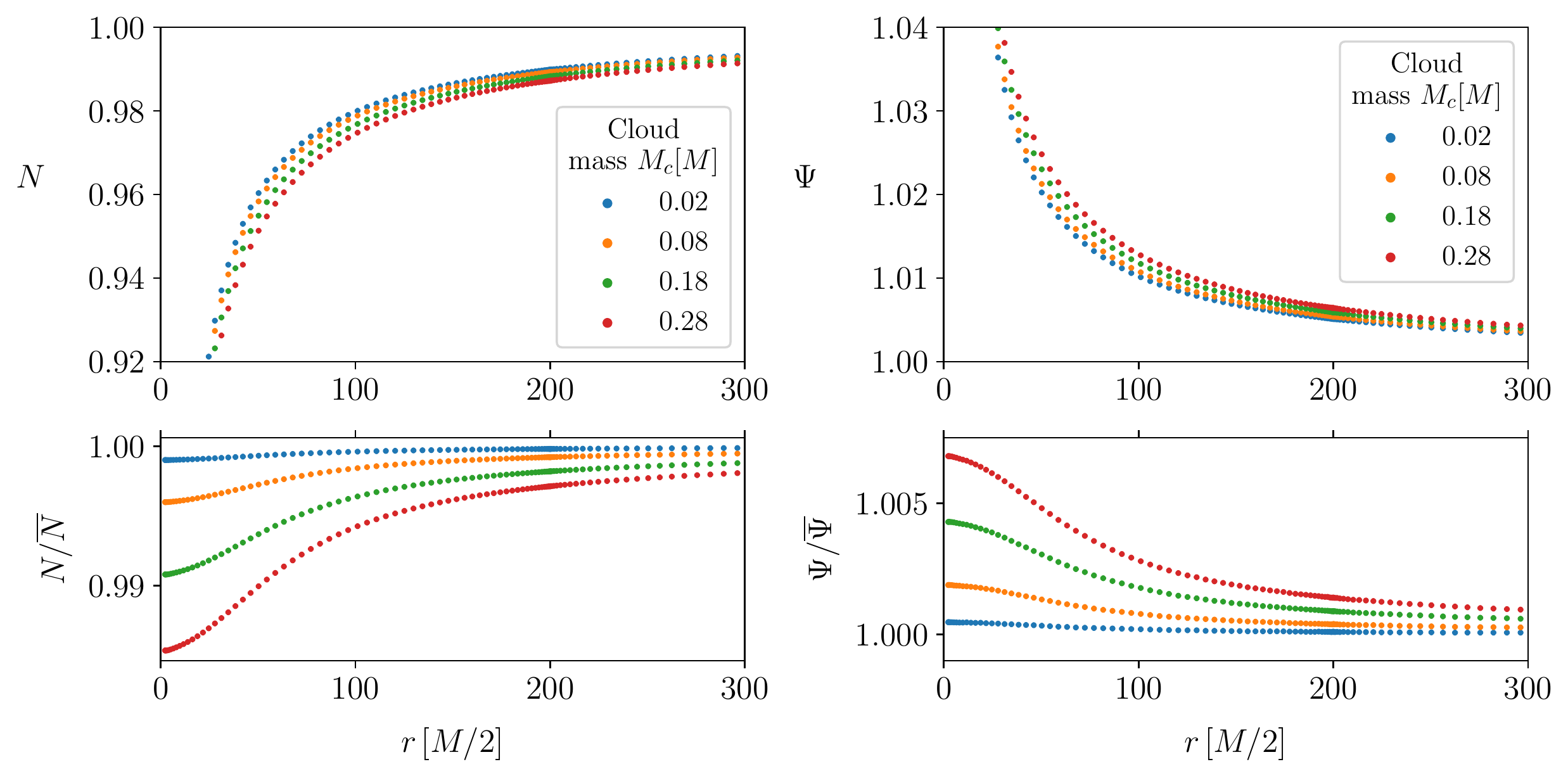}
    \caption{\textit{(left)}: The lapse function $N$, with the effects of backreaction of a boson cloud of $\alpha = 0.36$ and $\lambda = 0$ taken into account. The separate curves correspond to scenarios for different values of $M_{c}$ of the cloud, where $M_{c} \approx 0.29 M$ is the theoretical maximum efficiency of the Penrose extraction process~\cite{Penrose:1969pc}. \textit{(right)}: The equivalent results for the conformal factor $\Psi$. The bottom panels illustrate the ratio of the backreacted quantities to their Schwarzschild analogs shown in Fig.~\ref{fig:conflapse}.}
    \label{fig:br_N_Psi}
\end{figure}

Fig.~\ref{fig:br_N_Psi} shows the lapse function and conformal factor after backreaction from free scalar clouds. There we show the change of these metric quantities with increasing cloud mass $M_c$ (\ref{eqn:cloud_mass}), approximately up to the maximum mass of the cloud is determined by the theoretical maximum efficiency of the Penrose process, which is $M_{c} = ( 1 - 1/\sqrt{2} )M \approx 0.29 M$~\cite{Penrose:1969pc}. As $M_c$ increases, the redshift factor, $N$, monotonically decreases while the conformal factor monotonically increases, especially in the region between the black hole horizon and the Bohr radius of the cloud. This arises because the larger the $M_c$, the stronger the gravitational backreaction onto the spacetime, which results in a deeper the potential well and a redshift factor that vanishes near the horizon at a more rapid rate. Additionally, from the bottom panels of Fig.~\ref{fig:br_N_Psi}, we observe that the deviation between the backreacted and unbackreacted metric quantities follow a shape that resembles the radial profile of the cloud. Remarkably, Fig.~\ref{fig:br_N_Psi} also demonstrates that even in the strongest backreacting case, $M_c = 0.28 M$, the lapse and conformal factor only vary by approximately $1.5\%$ and  $0.7\%$ respectively near the horizon, compared to the unbackreacted case, cf. Fig.~\ref{fig:conflapse}. These small changes imply that backreaction effects on the spacetime curvature are overall marginal, with the Schwarzschild (and the more realistic Kerr) background spacetime serving as an excellent approximation for most practical purposes. 

\vskip 4pt

\begin{figure}[t!]
    \centering
    \includegraphics[width=\textwidth]{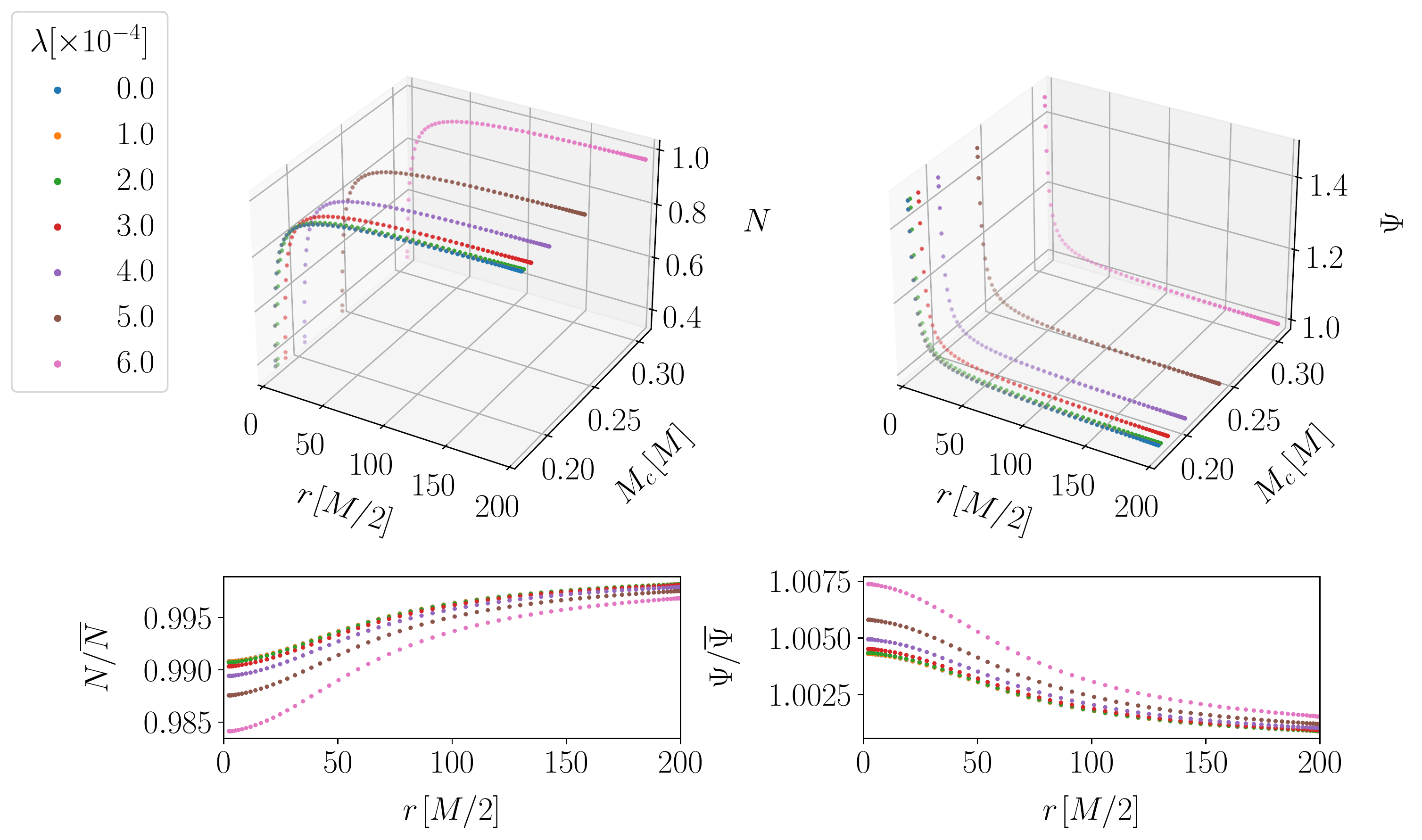}
    \caption{Same as Fig.~\ref{fig:br_N_Psi} except we include the nonlinearly backreacted $N$ and $\Psi$ for both free and self-interacting boson clouds. Here we fix $A=6 \times 10^{-4}$ and $\alpha=0.36$. The $\lambda=0$ curve here is identical to the $\lambda=0, M_c=0.18M$ curve in Fig.~\ref{fig:br_N_Psi}. Since $M_c$ depends on both $A$ and $\lambda$, we include a third axis in the top panel to illustrate the change of $M_c$ with $\lambda$. The results for $\lambda < 0$ are quantitatively and qualitatively similar to those shown here for $\lambda > 0$; see bottom panels of Fig.~\ref{fig:phi_lambda}.
    }%
    \label{fig:lapsepsi3d}%
\end{figure}

In Fig.~\ref{fig:lapsepsi3d} we display results which include the backreaction of self-interacting boson clouds on the lapse and conformal factor. In the upper 3D panels, we plot the radial profiles of $N$ and $\Psi$, and show an additional axis demonstrating the cloud mass $M_c$ of the boson cloud for a given value of $\lambda$. Note that we fix $\alpha=0.36$ and $A=6 \times 10^{-4}$ for all of those curves, such that the differences in $M_c$ arise entirely from contributions of the scalar field  potential. The range of values of $\lambda$ in Fig.~\ref{fig:lapsepsi3d} is chosen based on the monotonicity of the $M_c-\lambda$ relation discussed in \S\ref{sec:interacing_scalars}. In the lower panel of Fig.~\ref{fig:lapsepsi3d}, we show the ratio of $N$ and $\Psi$ to their values without backreaction. Similar to the free field case in Fig.~\ref{fig:br_N_Psi}, the changes in the nonlinear $N$ and $\Psi$ are rather marginal. Even for the largest explored value of $\lambda=6 \times 10^{-4}$, with the mass of the cloud reaching $M_c\approx 0.305M$, the change relative to the unbackreacted $\lambda=0$ case near the horizon is only $1.5\%$ and $0.75\%$ for the $N$ and $\Psi$ respectively. All in all, our analysis suggests that gravitational backreaction of the boson cloud onto the spacetime metric is rather marginal even in the strongest backreacting case.

\subsubsection{Parameter Space Exploration } \label{sec:paramspace}

In \S\ref{sec:SchwBackground} and \S\ref{sec:backreaction} we studied the radial distributions of $N, \Psi$ and $\Phi^N$ by focusing on a few choices of parameters to illustrate our key findings. In fact, we have performed a more comprehensive sampling in the $M_c - \lambda$ parameter space in order to understand behavior of the scalar field solutions and to test the regime of validity of our numerical methods. 

\vskip 4pt

\begin{figure}[h!]
\centering
\includegraphics[width=\linewidth]{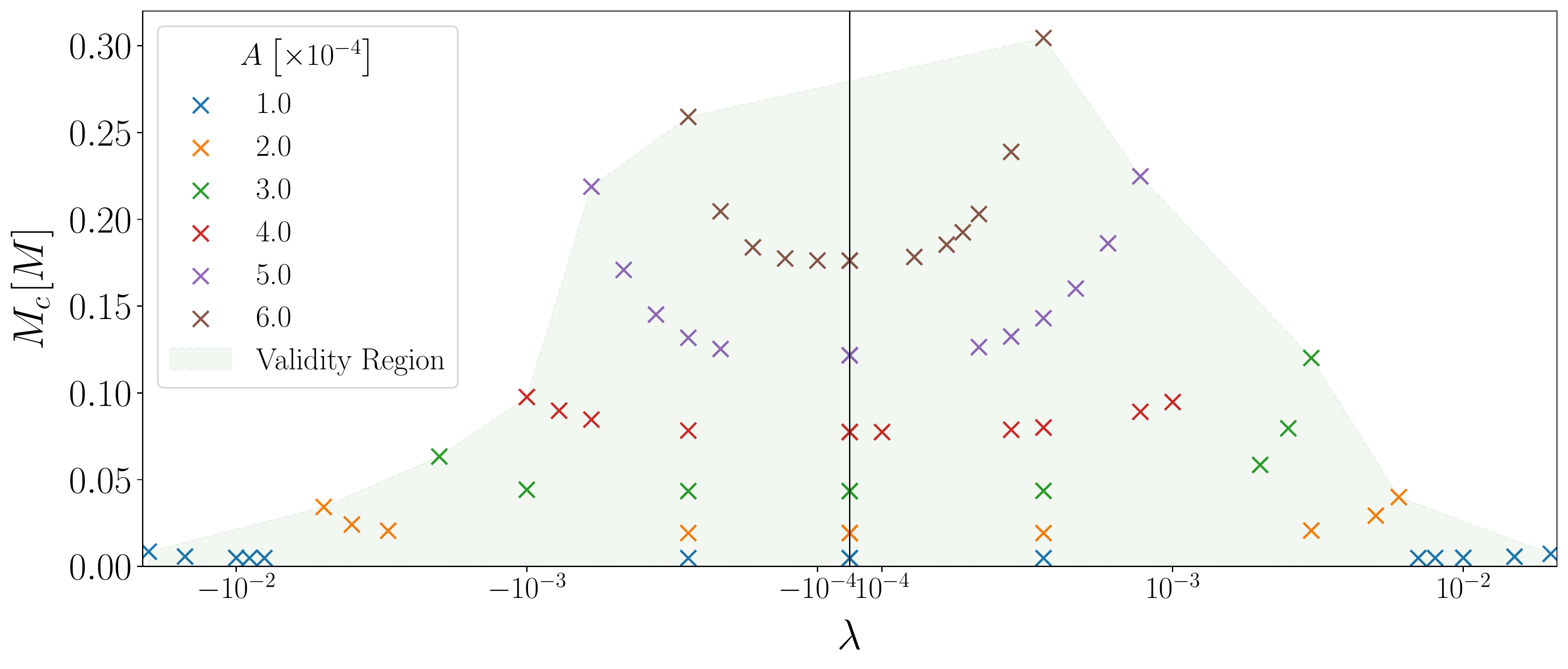}
\caption{A comprehensive exploration of the $M_c - \lambda$ parameter space with fixed $\alpha=0.36$. To produce this plot we sampled over a range of values for $\lambda$ and $A$, both of which determine $M_c$ in a nonlinear manner. The data points correspond to solutions with $M_c$ which change monotonically with $\lambda$. Numeric points outside of the shaded green region have $M_c$ which depart significantly from these monotonically-increasing trends and are not shown in this figure; see the bottom panels of Fig.~\ref{fig:phi_lambda} for an explicit illustration for $A=0.0006$. 
The shaded green region therefore represents the approximate region of parameter space where our numerical methods are robust.
}
\label{fig:paramspace}
\end{figure}

Fig.~\ref{fig:paramspace} shows an extensive exploration of the $M_c - \lambda$ parameter space, with the goal of determining the range of parameters that our method in Section~\ref{sec:setup} generates reliable solutions. The data points correspond to numeric results that follow the trend in which $M_c$ increases monotonically as the magnitude of $\lambda$ increases. Note that the values of $M_c$ are determined nonlinearly by $\lambda$ and $A$; in Fig.~\ref{fig:paramspace} we find that $M_c$ scales approximately quadratically in $\lambda$ (though there are also nontrivial couplings between $\lambda$ and $A$, see more discussion below). Outside of the shaded green region, the numeric data points have values of $M_c$ that, although convergent, depart significantly from these monotonically-varying trends; see the bottom panels of Fig.~\ref{fig:phi_lambda} for an explicit illustration for $A=0.0006$. This suggests that our method is no longer robust outside of the shaded green region and those solutions are physically unreliable. We suspect the main reason behind this ill behaviour lies in our use of the eigenfrequency of a free scalar field as an approximation to the eigenfrequency of the self-interacting scalar field, cf. \S\ref{sec:stationaryfield}, which is a key element in selecting the desired eigenstate in our numerical method. All in all, the green interpolated region illustrates an approximate region of parameter space where our numerical methods yield robust and reliable results. 

\vskip 4pt

\begin{figure}[t!]
    \centering
    \includegraphics[width=\textwidth]{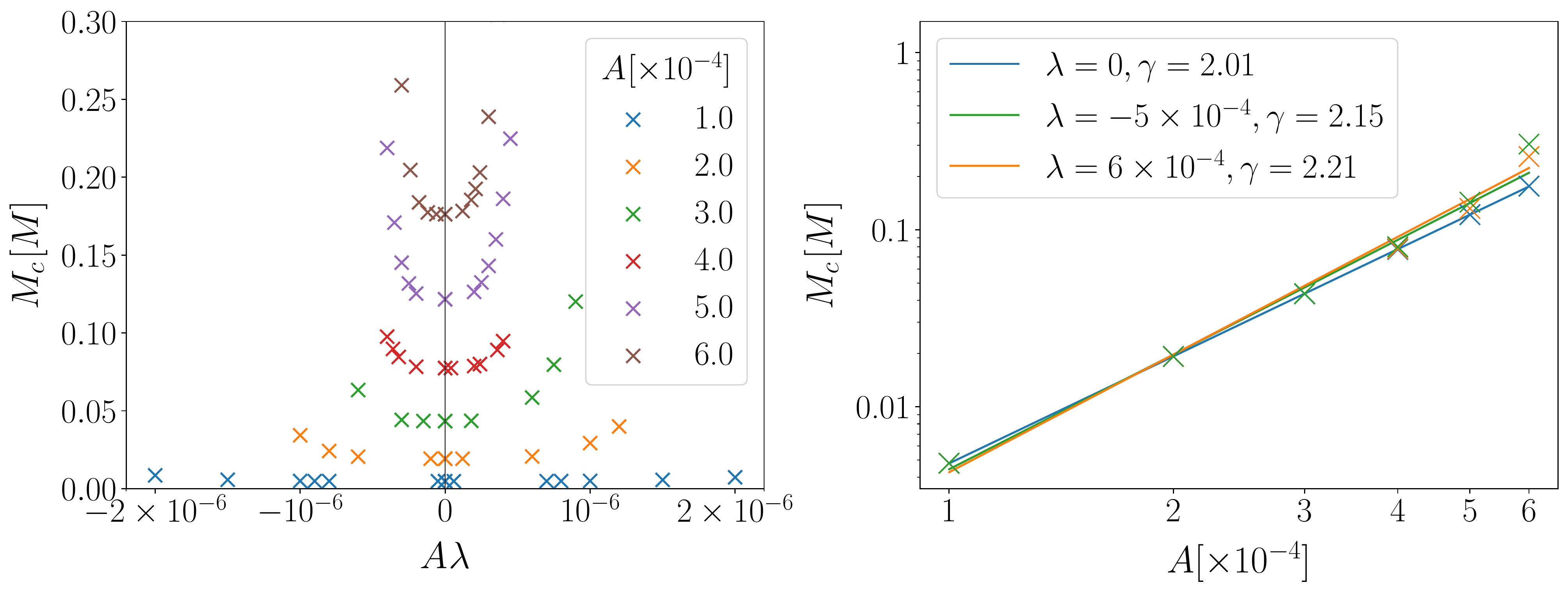}
\caption{\textit{(left)} Same as Fig.~\ref{fig:paramspace} except we plot $M_c$ along the $A\lambda$ axis and do not show the shaded region for simplicity. The non-trivial, approximately quadratic $M_c \sim (A \lambda)^2$, represents the nonlinear couplings between the scalar field inner boundary condition value $A$ (and hence the field amplitude) and the scalar interaction strength. \textit{(right)} The mass of the cloud $M_c$ as a function of the constant $A$ in log-log scale. The gradient of the straight line corresponds to the scaling parameter in the $M_c \propto A^\gamma$ ansatz, and are larger than the free-field case when scalar nonlinearities are present.
}
\label{fig:ADMamp}
\end{figure}

Since $M_c$ is a gauge invariant quantity, it is valuable to further explore its scalings with $A$ and $\lambda$ in our results. In the left panel of Fig.~\ref{fig:ADMamp}, we find that $M_c$ scales with the combination $A \lambda$ in a non-trivial manner, representing the nonlinear interactions in the scalar potential. In particular, we observe that $M_c$ scales approximately quadratically in $A \lambda$ in lines of constant $A$. This indicates that a fitting function for  $M_c(\lambda, A)$ would generally consist of a quartic polynomial in $A$ and $\lambda$, such that the combination $\sim (A \lambda)^2$ would appear as one of the terms in the polynomial function. The fact that a quadratic scaling, instead of linear scaling, in $\lambda A$ provides a better fit suggests that second-order and higher-order effects are already at play at these relatively modest values of $\lambda$ and $A$. 

\vskip 4pt

Finally, we explore the scaling of $M_c$ with $A$ for fixed values of $\lambda$. To this end, it is instructive to consider $M_c$ for the free scalar cloud, where the analytic expression is known. In the $\alpha \ll 1$ limit~\cite{Baumann2019, Brito2015a},
\begin{equation} \label{eqn:free_scalar_Mc}
     M_c (\lambda=0) = A^{\prime 2} \mu \propto A^2 \mu \, ,
\end{equation}
where $A^\prime \propto A$ is the amplitude of the scalar profile and is directly proportional to the value of the inner boundary condition we impose at $r_{\rm layer}$, cf. (\ref{eq:IB}). The free-field cloud mass (\ref{eqn:free_scalar_Mc}) has a simple interpretation: it is equal to the mass of the boson $\mu$ multiplied by the occupation number $A^{\prime 2}$ of the field configuration. Motivated by (\ref{eqn:free_scalar_Mc}), we use the ansatz for a self-interacting scalar cloud
\begin{equation}
   M_c (\lambda) \propto A^{\gamma(\lambda)} \mu \, ,
\end{equation}
where the scaling index $\gamma$ varies with $\lambda$ and satisfies $\gamma(\lambda=0) = 2$. In the log-log plot shown on the right panel of Fig.~\ref{fig:ADMamp}, in addition to confirming the $\gamma=2$ scaling for the free scalar field, we find that $\gamma$ is greater than two for both attractive and repulsive potentials. Qualitatively, the larger the magnitude of $\lambda$, the further the departure from the quadratic scaling. This finding agrees with the intuition that the the scalar potential contributes to the total energy budget contained in the boson cloud.

\pagebreak

\section{Conclusion and Outlook} \label{sec:conclusion}

Searches for dark matter often rely on the assumed presence of electromagnetic, weak or strong nuclear couplings between the dark matter and ordinary matter. However, if dark matter consists of ultralight bosons, they would be amplified spontaneously around rotating black holes through gravity, as demanded by the equivalence principle. A new window of opportunity to probe these ultralight fields with black holes have been made possible through the direct detection of gravitational waves~\cite{LIGOScientific:2014pky, VIRGO:2014yos}, either through nearly-monochromatic gravitational wave emissions from these clouds~\cite{Arvanitaki2011, Yoshino2014, Arvanitaki2015, Brito:2017wnc} or through their signatures imprinted on binary black hole waveforms~\cite{Chia:2020dye, Baumann2020a, Zhang:2018kib, Baumann2019a, Ding:2020bnl, Baumann:2021fkf}. When scalar self interactions are appreciable, the boson cloud could also emit scalar radiation~\cite{Arvanitaki2011, Yoshino:2012kn, Baryakhtar:2020gao, Omiya:2022gwu} and provide useful sources for axion dark matter detection experiments~\cite{Kahn:2016aff, JacksonKimball:2017elr, Chaudhuri:2019ntz, Lasenby:2019prg, Berlin:2020vrk}.

\vskip 4pt

In this paper we elucidated properties of self-interacting boson clouds around black holes in General Relativity, paying special attention to the effects of scalar self-interaction and the boson cloud's gravitational backreaction onto the spacetime metric. We used the spectral code \texttt{Kadath}~\cite{Grandclement2010} to solve the Einstein-Klein-Gordon field equations in the XCTS formalism~\cite{Pfeiffer2003, Gourgoulhon:2007ue}, which involves solving the Hamiltonian constraint, momentum constraint and the trace of the spatial evolution equation. The fact that a black hole resides at the center of the boson cloud presents new technical challenges to numerical implementations, for instance,  imposing the appropriate boundary condition at the black hole event horizon, which are not found in other well-studied relativistic scalar configurations such as boson stars. We described our approach to overcoming these challenges and discussed several approximations used which simplified our numerical evaluation of this complicated coupled set of equations in \S\ref{sec:Solve}. 

\vskip 4pt

In \S\ref{sec:SchwBackground}, we tested our method by evaluating the Schwarzschild background and the free scalar boson cloud solutions, finding excellent agreement between those numerical results and known analytical expressions. Our main results are presented in \S\ref{sec:backreaction}: for the self-interacting scalar fields with $- 10^{-2} \lesssim \lambda \lesssim 10^{-2}$, we find that the masses of the boson cloud are always larger than that of free scalar field. Although these self-interacting scalar cloud can have masses up to $\sim 70\%$ of that for a free scalar cloud, even reaching up to the $29 \%$ Penrose limit, the backreaction effects on the lapse and conformal factor are rather marginal due to the low-density nature of the clouds, cf. Figs.~\ref{fig:phi_lambda} and~\ref{fig:lapsepsi3d}. In addition, we solved the equations over a wide range in the $M_c-\lambda$ parameter space, as shown in Fig.~\ref{fig:paramspace}. For large values of $M_c$ and $A$ we find that the solutions beyond $- 5 \times 10^{-4} \lesssim \lambda  \lesssim 6\times 10^{-4}$ drastically depart from the expected monotonic change between $M_c$ and $\lambda$; see the lowest panel of Fig.~\ref{fig:phi_lambda}. For smaller values of $M_c$ and $A$, such monotonic behaviour extends further to the $- 10^{-2} \lesssim \lambda  \lesssim 10^{-2}$ regime. We suspect this departure indicates a sufficient difference between the eigenfrequencies of self-interacting and free-test-field configurations, which we use as an input in our setup, leading to convergent but seemingly unreliable outputs. Turning the argument around, this observation bounds the approximate region of parameter space where a self-interacting scalar eigenstate is adequately approximated by the free scalar field solution. Finally, we investigated the dependence of the cloud's mass on the self-interaction strength $\lambda$ and the scalar field's amplitude at the inner boundary $A$, finding approximate quadratic scaling relations $M_c$ with $A, \lambda$ and $A \lambda$ shown in Figs.~\ref{fig:paramspace} and~\ref{fig:ADMamp}. These results illustrate the nonlinear interactions induced by the scalar potential and shed new insights into the properties of these self-interacting scalar clouds.

\vskip 4pt

Our work is among the first systematic numerical explorations of self-interacting boson clouds in General Relativity and can be extended in several interesting directions: 

\begin{enumerate}

\item \textit{Self-interacting scalar field eigenfrequencies} -- We believe the main obstacle prohibiting us from a wider exploration in the range of $\lambda$ lies in our approximation of the scalar field eigenfrequency with the free-test-field eigenfrequency, cf. \S\ref{sec:stationaryfield}. Since $\omega$ is taken as an input in the XCTS system of equations, it has to be solved independently through other means. A brute-force grid scanning over various eigenfrequency values, for example as performed in~\cite{Fodor:2013lza}, would in principle overcome this problem but would be computationally prohibitive for our system. A spectral solver constructed such as those developed for the free scalar field in~\cite{Dolan:2007mj, Baumann2019} would certainly be useful towards achieving this goal, though unlike the free scalar case this would involve solving a nonlinear eigenvalue problem which is notoriously difficult~\cite{guttel2017};

\item \textit{Self-interacting scalar field eigenstates, or equivalently ``mode couplings"} -- Related to the previous point, our approximation of $\omega$ with the $\ket{211}$ dominant free-field eigenfrequency implicitly restricts ourselves to solutions that are weakly perturbed from the free field solution. In the language of free-field eigenstate basis, our setup is equivalent to having only weak couplings between $\ket{211}$ and other states~\cite{Yoshino:2012kn, Baryakhtar:2020gao, Omiya:2022gwu} are small. This also explains why our method is only robust in the rather limited range $-10^{-2} \lesssim \lambda \lesssim 10^{-2}$; for larger magnitudes of $\lambda$ we no longer expect the free-field eigenstates to be a adequate basis. In fact, when $|\lambda|$ is large the challenge of solving for these nonlinear eigenstates goes hand-in-hand with solving for the full eigenvalue problem~\cite{guttel2017} stated above;

\item \textit{Real scalar fields} -- We restricted ourselves to a complex scalar field because the $U(1)$ global symmetry ensures that the solution is stationary and axisymmetric, which simplify the system of equations. Although we do not expect our qualitative and quantitative results to differ substantially for those of a real field, it would be worth checking if this is true;  

\item \textit{Black hole spin} -- We have ignored the black hole angular momentum by imposing spherical symmetry in our setup, which enabled us to associate any departures from the free-field case to the scalar potential. The gravitational potential arising from the black hole spin is known to be subleading compared to the mass of the central black hole, for instance only leading to small ``hyperfine splittings" in the eigenfrequency spectrum of free scalar fields~\cite{Baumann2019, Baumann2019a}, and we do not expect the non-vanishing spin to significantly alter our results. Having said that, it would be interesting to explore how the solutions would change when we relax the $\beta^i = 0$ constraint to achieve an axially-symmetric spacetime, which would also enable the computation of the ADM angular momentum distribution of the system;

\item \textit{Spin-induced moments and tidal deformability} -- In addition to their masses and spins, it would also be interesting to compute the self-interacting boson clouds' spin-induced mass and current multipole moments and their tidal deformabilities in future work (see~\cite{Baumann2019a, DeLuca:2021ite} for results for free boson clouds). These observables are especially interesting when these clouds are in binary systems, as they are expected to introduce significant departures in the phases of gravitational waveforms from those of black holes~\cite{Hansen:1974zz, Thorne:1980ru, Binnington:2009bb, Damour:2009vw, Chia:2020yla, Charalambous:2021mea}, thereby affecting their detectability using standard binary black hole templates~\cite{Chia:2020psj, Chia:2022rwc}. 
    
\item \textit{Dynamical evolution} -- By solving the initial value problem instead of the full dynamical evolution of the system, our results only represent a quasi-equilibrium solution of the system and do not take dynamical effects, such as possible occupation of other free eigenstates during the instability, into account. Nevertheless, our results represent a good first approximation to the end state of the superradiance instability, and are in fact precisely the necessary initial data needed for its full numerical relativity evolution. We leave an investigation into its dynamical evolution in future work;

\item \textit{Singularity of the Klein-Gordon equation at the event horizon} -- We implemented a thin boundary at $r=r_{\rm layer}$ to impose the appropriate boundary condition and avoid the singularity of the Klein-Gordon equation at the event horizon, cf. \S\ref{eq:isolated_2nd_OD}. While we have tested and verified the effectiveness of this approach, it would be interesting to explore other methods that would more accurately implement the true only-ingoing condition at the black hole horizon, for instance by using a choice of coordinates that is regular at the event horizon~\cite{Grandclement:2022wif}

\end{enumerate}

We hope to address these questions in future work.

\subsection*{Acknowledgements}

We thank Philippe Grandcl\'ement for various enlightening discussions and for his patient guidance on using \lib{Kadath}~\cite{Grandclement2010}. We thank John Stout for sharing his code for the eigenfrequency solver for the free scalar field developed in~\cite{Baumann2019}. HSC gratefully acknowledges support from the Institute for Advanced Study and the Rubicon Fellowship awarded by the Netherlands Organisation for Scientific Research (NWO). AW and SN acknowledge support from the NWO Projectruimte grant (Samaya Nissanke). TH acknowledges the NWO sector plan and Cost Action CA18108.

\pagebreak

\newpage
\phantomsection
\addcontentsline{toc}{section}{References}
\bibliographystyle{utphys}
\bibliography{Numerics}
\end{document}